\documentclass{article}

\usepackage{PRIMEarxiv}

\usepackage[utf8]{inputenc} % allow utf-8 input
\usepackage[T1]{fontenc}    % use 8-bit T1 fonts
\usepackage{hyperref}       % hyperlinks
\usepackage{url}            % simple URL typesetting
\usepackage{booktabs}       % professional-quality tables
\usepackage{amsfonts}       % blackboard math symbols
\usepackage{nicefrac}       % compact symbols for 1/2, etc.
\usepackage{microtype}      % microtypography
\usepackage{lipsum}
\usepackage{fancyhdr}       % header
\usepackage{graphicx}       % graphics
\graphicspath{{media/}}     % organize your images and other figures under media/ folder
\usepackage{enumitem}
\usepackage{comment}

\newenvironment{tightitemize}{%
  \begin{itemize}[nosep,leftmargin=1.5em]%
}{%
  \end{itemize}%
}
\usepackage{listings}
\usepackage{amsmath,amssymb} % math environments & symbols
\usepackage{tikz}            % draw diagrams in LaTeX
\usetikzlibrary{arrows.meta,positioning,calc,bending} % arrows/positioning/math curves
\usepackage{float}           % force here placement [H]
\usepackage{tabularx}        % auto-width table columns
\usepackage{array}           % custom column types
\usepackage{caption}         % caption formatting
\usepackage{xcolor}          % colors (names/RGB/HTML)

\usepackage{float}
\usepackage{tabularx}
\usepackage{booktabs}

\newcommand{\ReaderDisclaimer}[2]{%
\begin{center}
\fbox{\begin{minipage}{0.97\linewidth}
\textbf{Reader-only disclaimer for #1:} #2
\end{minipage}}
\end{center}
}

\title{
Design principles for the Construction of a Benchmark Evaluating Security Operation Capabilities of Multi-agent AI Systems 
}

\author{
  Yicheng Cai, Mitchell John DeStefano, Guodong Dong, Pulkit Handa, \\ \textbf{Peng Liu\thanks{Corresponding author}, Tejas Singhal, Peiyu Tseng, Winston Jen White} \\
  Cyber Security Laboratory \\
  Pennsylvania State University, University Park \\
  \texttt{\{yccai, mjd7100, gvd5289, pzh5320\}@psu.edu} \\
  \texttt{\{pxl20, tejas.singhal, pmt5342, wjw5326\}@psu.edu}
  }
\begin{document}
\maketitle

\begin{abstract}
As Large Language Models (LLMs) and multi-agent AI systems are demonstrating increasing potential in cybersecurity operations, organizations, policymakers, model providers, and researchers in the AI and cybersecurity communities are interested in quantifying the capabilities of such AI systems to achieve more autonomous SOCs (security operation centers) and reduce manual effort. In particular, the AI and cybersecurity communities have recently developed several benchmarks for evaluating the red team capabilities of multi-agent AI systems. However, because the operations in SOCs are dominated by blue team operations, the capabilities of AI systems \& agents to achieve more autonomous SOCs cannot be evaluated without a benchmark focused on blue team operations. To our best knowledge, no systematic benchmark for evaluating coordinated multi-task blue team AI has been proposed in the literature. Existing blue team benchmarks focus on a particular task. The goal of this work is to develop a set of design principles for the construction of a benchmark, which is denoted as {\bf SOC-bench}, to evaluate the blue team capabilities of AI. Following these design principles, we have developed a conceptual design of SOC-bench, which consists of a family of five blue team tasks in the context of large-scale ransomware attack incident response.  
\end{abstract}

% keywords can be removed

\section{Introduction} 

Security Operation Centers (SOCs) are addressing a fundamental societal challenge: enabling organizations (e.g., enterprises across various industry sectors, educational institutions, governments) to deal with cyber threats. 
The action-taking aspect of a SOC is supported by a sophisticated decision-making process, and this process involves a combination of both lower-order cognitive skills (e.g., summarizing, retrieval, recall) and higher-order cognitive skills (e.g., explanation, evaluation, reasoning). Accordingly, today's {\em cybersecurity operations} are heavily relying on manual effort, though security teams do leverage a variety of information technology tools such as Intrusion Detection systems, SIEM (Security Information Event Management) systems, and  SOAR (Security Orchestration, Automation and Response) tool suites. 

As Large Language Models (LLMs) and multi-agent AI systems are demonstrating increasing potential in cybersecurity operations, organizations,  
policymakers, model providers, and researchers in the AI and cybersecurity communities are interested in quantifying the capabilities of such AI systems to achieve more autonomous 
SOCs and reduce manual effort. 
In particular, the AI and cybersecurity communities have recently developed several benchmarks (e.g., Cybench \cite{zhang2025cybench}, CyberGym \cite{wang2026cybergym}) for evaluating the {\em red team} capabilities (e.g., penetration testing) of multi-agent AI systems. Taking Cybench \cite{zhang2025cybench} as an example, it includes 40 professional-level Capture the Flag (CTF) tasks from 4 distinct CTF competitions, spanning a wide range of difficulties.

However, because the operations in a SOC are dominated by {\em blue team} operations, the capabilities of multi-agent AI systems to achieve more autonomous SOCs cannot be evaluated without a benchmark focused on blue team operations. (To the best of our knowledge, no systematic benchmark for evaluating coordinated {\bf multi-task} blue team AI  
has been proposed in the literature. Existing blue team benchmarks focus on a particular task (e.g., indicators extraction from threat reports). )  
The goal of this work is to develop a set of design principles for the construction of a benchmark, which is denoted as {\bf SOC-bench}, to evaluate blue team operation capabilities of multi-agent AI systems. 

Due to the fundamental differences between red team and blue team operations, the design principles we seek to develop are very different from the existing work focused on developing a benchmark evaluating read team AI.  In particular, five design principles are identified: (DP1) Except for only allowing humans to serve as a supervisor, the as-is SOC (incident response) system, 
not the to-be SOC system, serves as the gold standard. 
(DP2) Because AI systems should autonomously explore and leverage the interdependence between the
family of blue team tasks, SOC-bench should not provide AI systems with any cross-task hints. 
(DP3) AI systems could go beyond mimicking humans.
(DP4) Real-world SOC (incident response) systems are inherently imperfect.
(DP5) SOC-bench should not be specific to any current capabilities of AI systems. 
In order to validate these design principles, we have followed these principles and developed a conceptual design of SOC-bench which consists of a family of five blue team tasks in the context of large-scale ransomware attack incident response: (Task Fox) Early cyberattack campaign detection; (task Goat) file system forensics; (task Mouse) data exfiltration analysis; (task Tiger) analysis of indicators of compromise (IOCs) to attribute the attack and report the used Tactics, Techniques, and Procedures (TTPs); (task Panda) recommending containment actions.   

%which consists of five XXX. 

The rest of this paper is organized as follows. In Section 2, we explain why a benchmark focused on evaluating blue team AI agents would play an indispensable role in the technology adoption lifecycle of real-world SOCs in the era of AI. In Section 3, we present a set of design principles for SOC-bench. In Section 4, we provide an overview of the {\em conceptual design} of SOC-bench. From Section 5 to Section 9, we present the five major blue team tasks of our conceptual design. In Section 10, we conclude the paper.      

%--------------------------
\section{Why SOC-bench?}

SOC-bench is clearly not the first benchmark in the intersection of Cybersecurity and AI. 
In fact, there are a number of benchmarks (e.g., \cite{wu2025excytinbenchevaluatingllmagents, jajodia2026handwheelevaluatingllms,metacyberseceval4}) in the intersection. 
From the viewpoint of real-world cybersecurity operations, the existing benchmarks as well as
the potentially meaningful benchmarks can be broken into the following categories: 

\begin{itemize}
    \item {\bf Red Team - Penetration Testing benchmarks}. Benchmarks (e.g., \cite{zhu2025cvebenchbenchmarkaiagents}) in this category assume the standard real-world penetration testing setup, where each to-be-tested vulnerability has a particular CVE number.    
    
    \item {\bf Red Team - Vulnerability Discovery benchmarks}. Benchmarks (e.g., \cite{zhang2025bountybench}) in this category are used to evaluate the capabilities of AI systems to identify unknown vulnerabilities.  
    
    \item {\bf Red Team - Competition}. Benchmarks in this category are used to evaluate the capabilities of AI systems to win a competition. Such competition not only include Capture-the-Flag (CTF) competitions (e.g., the DARPA AI Cyber Challenge \cite{darpaaicyber}), but also include AI model specific competitions (e.g., adversarial examples against DNN models, backdoor attacks against AI models). 
    
    \item {\bf Blue Team - Single Task}. Benchmarks (e.g., \cite{froudakis2025revealingtrueindicatorsunderstanding}) in this category focus on a particular task performed by blue teams. For example, a benchmark may be developed to evaluate whether an AI system can effectively extract IOCs from CTI (Cyber Threat Intelligence) reports. For another example, a benchmark can be developed to evaluate whether an AI system can correctly generate patches.  

    \item {\bf Blue Team - Incident Response}. In the presence of large-scale cyberattack campaigns, real-world incident response processes are never a single task process. In contrast, it is a sophisticated real-time process that involves seamless coordination of multiple security teams working on a family of complementary tasks. Some of the representative tasks include early detection of cyberattack campaigns, preliminary system forensics, data exfiltration analysis, attack containment, threat intelligence analysis, risk management, and recovery.  

    \item {\bf Blue Team - Competition}. Benchmarks (e.g., \cite{kiely2023autonomous}) in this category are used to evaluate the capabilities of AI systems to win an defense-centric competition. Being a competition, each benchmark in this category usually deviates from real cybersecurity operations to certain extent.    
\end{itemize}

SOC-bench belongs to the Blue Team - Incident Response category. 
Regarding why there is clearly a need to develop SOC-bench, our main reasons are as follows. 
First, because real-world blue team operations are {\em dominated} by 
real-time incident response processes, we argue that a benchmark which is 
most suitable for evaluating the {\em core} capabilities of 
of a Security Operation Center should be a benchmark in the  
Incident Response category. 

Second, no existing benchmark belongs to the Blue Team - Incident Response category. 
Although there are already a few benchmarks in the category of Blue Team - Single Task, 
the union of these benchmarks won't become a meaningful benchmark in 
the Incident Response category. Since these benchmarks are independently 
developed, the connections (e.g., dependencies) between the individual tasks handled by these
benchmarks significantly deviate from those between 
the individual tasks conducted in one real incident response process. 

Third, we observe that the cybersecurity operation industry not only needs 
one or more benchmarks in the 
Blue Team - Incident Response category, but also needs  
the benchmarks to cover both the current AI adoption best practices  
and the future evolutions of AI. 
With respect to SOCs, on the one hand, the current AI adoption best 
practices seem to be treating AI models (e.g., LLMs) and AI agents as tools utilized by 
human analysts. 
On the other hand, it is increasingly recognized that the rapidly 
   emerging {\em agentic AI} technology (e.g., OpenClaw \cite{openclawgit}) could soon make SOCs substantially more {\em autonomous}. 
Accordingly, a benchmark which focuses on the incident response processes 
  where humans only serve as a supervisor and multi-agent AI systems do the whole job 
  would be increasingly relevant to real-world cybersecurity operations. 
This is exactly what SOC-bench aims to achieve.

%        \item The scope of SOCBench cannot be taken for guranteed
%        \item Why the midway on the spectrum is esential for development of agentic AI in cybersecurity

%============================================
\section{Design Principles of SOC-bench}

When developing SOC-bench, we have identified the following five design principles. 

\begin{enumerate}
    \item {\bf DP1.} Except for only allowing humans to serve as a supervisor, the {\bf as-is} SOC (incident response) system, 
    NOT the to-be SOC system, serves as the gold standard. The implications of the golden standard include but are not limited to the following: (a) SOC-bench should be designed based on a large scale cyberattack campaign that had already happened in 
    the real world. No future attack should be considered.  
    (b) Every data source included in SOC-bench should be a commonly used type of data source. 
    (c) AI systems don’t have any extra knowledge than human security teams regarding what the adversary 
    knows about the target network and the cyberattack campaign. The entire SOC-bench should be based on 
    what a SOC observes and knows. 
    (d) The amount of uncertainty being addressed by human teams will not change due to deployment of AI systems. 

    \item {\bf DP2.}  Because AI systems should autonomously explore and leverage the interdependence between the family of blue team tasks, SOC-bench should not provide AI systems with any cross-task hints. For example, SOC-bench should not provide any hint regarding the data or control dependencies between any two tasks. For another example, SOC-bench should not provide any hint regarding which kinds of multi-agent coordination are beneficial.

    \item {\bf DP3.} AI systems could go beyond mimicking humans. Accordingly, first, the extent to which the problem-solving procedures (including all the detailed action-taking) of AI systems deviate from those of human security teams should not be taken into consideration when evaluating the blue team capabilities of AI systems. Second, the outcome-only principle should be followed when SOC-bench is used to evaluate the blue team capabilities of AI systems. That is, given that the input to every AI system is always the same, SOC-bench should explicitly specify the specific required outcomes. In addition, for each required outcome, the rating system must not take any problem-solving procedures into consideration. 

    \item {\bf DP4.} Real-world SOC (incident response) systems are inherently imperfect. For example, 
    IDS alerts and SIEM alerts  
          are imperfect, suffering from false positives, false negatives, and latency. 
    For another example, in many cases, not all the useful data sources (e.g., measurements of CPU energy consumption, SSD flash translation layer activities) are available. 

    \item {\bf DP5.} SOC-bench should not be specific to any current capabilities of AI systems. Agentic AI technologies are rapidly evolving. As a result, some currently popular techniques (e.g., the Model Context Protocol (MCP)) may or may not remain essential in the future. 
    In order to be future-proof, the design of SOC-bench should not be specific to particular current capabilities (e.g., RAG (retrieval augmented generation), MoE (mixture of experts)) of AI systems.  

\end{enumerate}

% ---------- start Overview ----------
\section{Overview of SOC-bench}
%\subsection{Task Overview}

%\paragraph{Objective:} SOC-bench is a benchmark framework developed to evaluate AI agent capabilities across the full ransomware incident lifecycle in a SOC environment. The goal is to test not just detection and containment but also awareness, reasoning, and decision-making under realistic constraints. SOC-bench does not create new tasks but replicates real SOC tasks and data from actual incidents. This approach ensures that the evaluation focuses on how AI agents would perform under the same operational conditions faced by real SOC analysts.

\subsection{Terminology}\label{sec:terms}

\begin{itemize}[leftmargin=1.25em]
%  \item \textbf{SOC-bench:} Benchmark framework that evaluates AI agents across the ransomware incident lifecycle in a SOC (Security Operations Center) environment.
  
  \item \textbf{Agentic AI System:} An AI system composed of multiple coordinated agents. Each agent is fulfilling a particular task.  
  %that can reason autonomously and select actions across SOC tasks.

  \item \textbf{AI Agent:} When fulfilling a task, an autonomous AI agent not only executes pre-planned actions, but also may conduct on-the-fly action planning.  
  %The AI system under evaluation that reasons from SOC-observable data and produces recommended actions or reports. 
  
  \item \textbf{Task:} Each task represents one of the blue team operation challenges (e.g., early cyberattack campaign detection).
  %(Fox, Tiger, Panda, Goat, Mouse) representing distinct SOC functions. 
  
  \item \textbf{Telemetry:} Logs, alerts, network flow data, events on a machine, and other data sources collected by SOC systems. 
  \item \textbf{Visibility Constraints:} Limits on the telemetry or data available to an agent during a task.
  \item \textbf{Ground Truth:} The facts of a cyberattack campaign obtained from post-incident analysis (not available to agents during blue team operations). 
  \item \textbf{IOC (Indicator of Compromise):} An information item such as a command line command, a registry key, a hash, an IP address, or a file path that signals potential malicious activity. 
  \item \textbf{CTI (Cyber Threat Intelligence):} Out-of-band information about adversary behavior, tools, or indicators. 
  \item \textbf{TTP (Tactics, Techniques, and Procedures):} These terminologies are defined by MITRE ATT\&CK framework \cite{mitre_attack_data_sources}. 
  %patterns, methods, or behaviors used by an attacker.
  \item \textbf{Inter-Task Dependencies:} The data and/or control dependencies across two or more blue team tasks. 
  %SOC functions that create natural ambiguity, requiring agents to reason carefully to avoid incorrect conclusions.
  \item \textbf{Containment Action:} A category of defense actions (e.g., host quarantine, network segmentation, process termination) aiming to contain a cyberattack campaign. 
  %intended to limit spread. 
  \item \textbf{VSS Snapshot:} Windows Volume Shadow Copy snapshot used to preserve prior file states.
  
  \item \textbf{SIEM (Security Information and Event Management):} This terminology refers to real-world SIEM systems. 
  %A system that collects and correlates security alerts, logs, and telemetry from across the network to support detection and incident analysis.
  
  \item \textbf{Lateral Movement:} The technique used by attackers to move from an initial foothold to additional systems within a network. 
\end{itemize}

\subsection{Scope}
%\textbf{Blue Team Perspective:} 
(a) Following design principle DP1, SOC-bench represents one real-world as-is SOC incident response system during 
a particular time window. During the time window, a large-scale ransomware attack incident happened. 
(b) All other time windows of the SOC incident response system are out of the scope of SOC-bench. 
(c) All other kinds of cyberattack campaigns are out of the scope. 
(d) All red team operations are out of the scope. 

%The benchmark reflects what a SOC sees during . Inputs come from the defender side telemetry, logs, alerts, and employee reports. Attacker-side details or hindsight information are not included.

\subsection{Network Topology}
The SOC-bench network mirrors the enterprise topology observed during the Colonial Pipeline ransomware incident \cite{ColonialWiki}.  It mirrors the same segmentation weaknesses, host types, and network traffic patterns that shaped the SOC’s actual response. Critical systems such as file servers, Active Directory controllers, VPN servers, and operational technology (OT) devices remain connected through limited segmentation, allowing partial lateral movement between subnets. 
Each subnet represents a functional area of the enterprise, including corporate IT, OT, and remote access infrastructure. This structure allows AI agents to reason about containment and propagation under the same situation 
%within the same boundaries 
faced by human SOC analysts. 

Following design principle DP1, the network topology is not redesigned or optimized but faithfully derived from post-incident data and telemetry to ensure authenticity. The network context serves as the foundation for all SOC-bench tasks. It allows evaluations to be conducted under realistic visibility constraints, ensuring that AI agents operate under the same physical and logical conditions as the original SOC investigation.

\subsection{Ransomware Characteristics in Real-World Incidents} 
\paragraph{Attack Initiation:} Most ransomware attacks, including the Colonial Pipeline ransomware, begin through a compromised access point, such as a VPN exploit. 
This aligns with Task Tiger.  
%Task Fox (Early Attack Detection), which evaluates whether agents can identify coordinated ransomware activity early but not prematurely based on real SOC data.
\paragraph{Lateral Movement:} After initial access, attackers move laterally toward critical systems such as file servers and AD controllers. Because SOC-bench reflects a network with limited segmentation, AI agents must reason about containment under realistic restrictions. 
This stage aligns with Task Fox and Task Panda. 
%(Containment), which measures whether agents can detect spread and suggest isolation actions. 
\paragraph{Impact:} Once critical systems are reached, the ransomware encrypts files and disrupts operations. This phase aligns with Task Goat and Task Mouse.  
%, in which agents analyze metadata and logs to determine which files are encrypted, identify the responsible processes, and gauge the impact.

%\subsection{Threat Context and Data Complexity}
%\paragraph{Ransomware Basis:} SOC-bench tasks model ransomware behaviors observed in incidents such as Colonial Pipeline and DarkSide. 
%\paragraph{Incident Reconstruction:} Multi-stage attack sequences are reconstructed from SOC-observed data, including lateral movement, privilege escalation, and decoy activity. Each reconstructed attack sequence reflects realistic SOC operations and is informed by incident telemetry and post-incident analysis. The benchmark is intentionally difficult, requiring agents to reason under uncertainty and partial information. This approach evaluates AI performance under similar operational conditions faced by real SOC analysts, rather than relying on artificial or synthetic environments. Scores are expected to be low at first, reflecting the complexity of genuine SOC workflows rather than deficiencies in model capability.
%\paragraph{Data Integration:} Data sources are unified, so agents must connect events across layers such as network, endpoint, and SIEM.
%\paragraph{Detection Focus:} Agents are rewarded for minimizing false positives and reasoning through uncertain evidence rather than matching static signatures.

\subsection{Why SOC-bench Can Focus on the Five Tasks?}

Regarding why SOC-bench can focus on five aforementioned tasks, the main reasons are as follows. 

\begin{itemize}[leftmargin=1.25em]
  \item \textbf{Structure:} SOC-bench employs a five-task structure (Fox, Tiger, Panda, Goat, and Mouse). 
  \begin{itemize}
  \item {\em Task Fox:} Early cyberattack campaign detection. 
  \item {\em Task Goat:} File system forensics.
  \item {\em Task Mouse:} Data exfiltration analysis.
  \item {\em Task Tiger:} Analysis of IOCs to attribute the attack and report the used TTPs. 
  %Tactics, Techniques, and Procedures (TTPs); 
  \item {\em Task Panda:} Recommending containment actions.   
  \end{itemize}

  \item When responding to large-scale ransomware attacks, the blue team operations indeed include each of the five tasks. We argue that all of the five tasks are essential incident response tasks when defending against large-scale ransomware attacks.     

  \item We observe that these five tasks inherently cover a significant portion of the coordination activities among the blue team members. 

  \item We observe that these five tasks capture representative inter-task data and control dependencies. Since the captured dependencies are representative, if a multi-agent AI system can achieve remarkable performance against SOC-bench, it should be able to achieve remarkable performance when conducting real-world blue team operations.  
  %aligned with how a SOC processes ransomware cases from early detection to recovery. 
  
  \item \textbf{Reliable Evaluation:} We found that the performance of a multi-agent AI system against each of the five tasks can be evaluated in an objective way. 
  %Scoring emphasizes analytic depth, situational awareness, reasoning quality, and evidence alignment over simple correctness.
  %\item \textbf{Goal:} Encourage the development of AI systems capable of autonomous and explainable decision-making under incomplete information.
\end{itemize}

\section{Task Fox}\label{sec:fox}

\subsection{Task Overview}\label{sec:fox-overview}
%\textit{(Principles: DP1, DP2, DP3, DP4, DP5, No attacker loyalty)}

\paragraph{Objective:}
Task FOX evaluates a multi-agent AI system 
%(Agentic AI system coordinated through AI orchestration) 
based on the earliest feasible detection of a major coordinated attack campaign, without assuming the campaign type (ransomware or otherwise).

The unit checks the ability of the agentic AI system to:
\begin{itemize}[leftmargin=1.25em] 
  \item Determine whether the observed multi-host activity represents a major coordinated attack campaign or is it just ordinary isolated anomalies.
  \item If major, identify whether its behavioral signature (patterns etc.) is consistent with ransomware precursors (pre-T1486 activity) or something else entirely.
  \item Produce evidence backed claims under uncertainty (even when the data is incomplete), without relying on attacker disclosures. (Attacker=Lier)
  \item Operate solely based on SOC-visible observations and procedures, not ground-truth attacker behavior. (We do not know what the attacker does)
  \item Summarize all this situational awareness in a one-page report.
\end{itemize}

\paragraph{What the AI system must deliver.}
The AI system must produce three outcomes (more details later on):
\begin{enumerate}[leftmargin=1.25em]
  \item O1: Campaign-Scale Assessment: classification of activity as \{major campaign, small-scale anomaly, unknown\}.
  \item O2: Type Inference (Conditional on O1): if major, classification of \{ransomware-like, non-ransomware coordinated, unknown\}.
  \item O3: Incident Alert Triage and Summary: consolidated alert containing the decision, a list of signals (like multi-host anomalies, tool misuse indicators) with corresponding evidence, and a three-part summary at the end of all this.
\end{enumerate}

\textit{Reader-only note.} Following design principle DP3, this task specifies \emph{outcomes only}. Methods are deliberately unspecified; evaluation is by matching outcomes to ground truth via the scoreboard.

\subsection{Scope and Assumptions}\label{sec:fox-scope}
%\textit{(Principles: DP1, DP4, DP5)}
\begin{itemize}[leftmargin=1.25em] 
  \item Even if there are clear indicators of a large-scale cyberattack campaign, %(base attack taken), 
    FOX does not blindly assume ransomware attack.  
  \item Agents observe only SOC-visible telemetry: no attack campaign narrative is available. 
  \item Data logs are available in incomplete form.
  \item The benchmark does not score the real-time performance of the AI system; the performance evaluation is conducted per-time-window. %uses checkpointed stages.
  \item Agents may behave as fast detectors or lazy detectors (e.g, do not raise alerts until receiving user file failed-access reports). Both strategies are evaluated using the same scoring criteria. 
  %(Lazy maybe more trustworthy)
\end{itemize}

\subsection{Data Sources}\label{sec:fox-data}
%\textit{(Principles: DP1, DP4)}

\subsubsection{Host-Level Telemetry}\label{sec:fox-data-host}

%The following host-side measurements are often available to a SOC: 
\begin{itemize}[leftmargin=1.25em]
  \item Windows event logs. 
  \item Linux Auditing Subsystem logs. 
  \item EDR (endpoint detection and response) logs. 
  \item Application-specific logs. 
  \item System call trace on Linux machines. %for the log of High I/O bursts. 
  \item CPU and memory usage measurements. 
%  \item High I/O bursts (Input/Output) (Task Panda)
%  \item Unexpected process launches
%  \item Privilege-escalation attempts Can have in 4.4
%  \item Authentication spikes (multiple failed logons across hosts)
\end{itemize}

%\fbox{%
%  \parbox{\textwidth}{%
%    \textbf{Reader-only examples (not shown to agents):} process anomalies, abnormal resource consumption patterns etc.
%  }%
%}

%\subsubsection{Process Execution \& Administrative Tool Telemetry}\label{sec:fox-data-proc}

\begin{comment}

Process-level activity that is collected from standard enterprise logging sources such as Windows Event Logs, EDR process logs, and command-line auditing:
\begin{itemize}[leftmargin=1.25em]
  \item Process command-line activity
  \item Parent–child process chains
  \item Unexpected PowerShell execution on multiple hosts
  \item Remote administration tools such as PsExec service creation
  \item Service installation logs (e.g., Event ID 7045)
\end{itemize} 

\fbox{%
  \parbox{\textwidth}{%
    \textbf{Reader-only examples:} PowerShell running at the same time on several hosts, suspicious PsExec usage, Admin tools being used in places where they are normally not used etc.
  }%
}

\end{comment}

\subsubsection{Network Activity \& Lateral Movement Signals}\label{sec:fox-data-net}

\begin{itemize}
\item Firewall logs. 
\item PCAP files. 
\item Windows security logs.
\item NetFlow records (i.e., record of network conversation metadata). 
\end{itemize}

\begin{comment}

Network-level information available from firewall logs, Windows security logs, NetFlow records (which is record of network conversation metadata):
\begin{itemize}[leftmargin=1.25em]
  \item Inter-host authentication logs (Kerberos/NTLM activity)
  \item SMB/Share access attempts
  \item Network scanning detected in firewall or NetFlow logs
  \item Abnormal inbound or outbound connections through PCAP
\end{itemize} 

\fbox{%
  \parbox{\textwidth}{%
    \textbf{Reader-only examples:} Repeated authentication attempts from unexpected hosts, Internal scanning patterns that do not match normal operations across multiple subnets.
  }%
}

\end{comment} 

\subsubsection{SIEM Alert Stream}\label{sec:fox-data-siem}

%High-level alerts derived from host and network logs:
This data source holds the stream of SIEM alerts.  Following design principle DP4,
it should be noticed that SIEM alerts may be delayed and could be a false positive, and 
that SIEM systems can be affected by false negatives. 

\begin{comment}

\begin{itemize}[leftmargin=1.25em]
  \item Used only for supporting evidence
  \item Never replaces raw evidence
  \item Alerts are sometimes delayed or incomplete (normal in real SOCs)
  \item Not sufficient on their own for O1 or O2 decisions
\end{itemize}

\fbox{%
  \parbox{\textwidth}{%
    \textbf{Reader-only remark:} \\ SIEM alerts help confirm suspicious behavior but 
    should never be treated as enough proof by themselves alone.
  }%
}

\end{comment}

\subsubsection{Helpdesk Tickets}\label{sec:fox-data-helpdesk}

This data source holds user-reported issues. Some of the 
ticketed user messages are as follows. 
Note that these messages usually arrive later than telemetry. 
%Plain, non-technical messages from employees:
\begin{itemize}[leftmargin=1.25em]
  \item ``I cannot access my files.''
  \item ``Shared drive is not opening.''
  \item ``My applications are freezing.''
\end{itemize}

%-------------------------
\subsection{Outcomes}\label{sec:fox-outcomes}
%\textit{(Principles: DP1, DP3)}

Task FOX outputs are stage-scored. At each stage (e.g., 30-minute time interval), the AI system must emit three early detection outcomes.  
%that reflect what a SOC can observe from logs/telemetry (not attacker intent). 
Each outcome must cite concrete evidence objects (e.g., particular log entries). 

\paragraph{O1. Campaign-Scale Assessment (stage-level)}
\begin{itemize}[leftmargin=1.25em]
  \item \textbf{Goal:} Determine whether the suspicious activities in the current stage and the previous stages are \\
  (a) isolated/localized or \\
  (b) indicative of a campaign-scale cyberattack spanning multiple hosts.
  \item \textbf{Submit (JSON):} one JSON object o1\_scale containing:
\end{itemize}

\begin{verbatim}
{
    "o1_scale": {
        "stage_id": "T+60",
        "scale_label": "isolated | localized | campaign_scale",
        "impacted_hosts": ["HOST-A", "HOST-B"],
        "evidence_ids": ["EVID-123", "EVID-219"],
        "rationale": "1-2 sentences tying label to evidence"
    }
}
\end{verbatim}

Interpretation guidance (grader only section):
\begin{itemize}[leftmargin=1.25em]
  \item Isolated: one host, no cross-host correlation.
  \item Localized: multiple signals but confined to a single host or a single adjacency (e.g., one host + one server).
  \item Campaign-scale: clear multi-host correlation (e.g., authentication bursts across hosts, simultaneous anomalies, repeated remote-execution artifacts).
\end{itemize}

\paragraph{O2. Activity-Type Reasoning (conditional on O1)}
\begin{itemize}[leftmargin=1.25em]
  \item \textbf{Goal:} If O1 indicates campaign\_scale, decide whether behavior is ransomware-like staging/execution vs non-ransom coordinated activity.
  \item \textbf{Submit (JSON):} one JSON object o2\_type containing:
\end{itemize}

\begin{verbatim}
{
    "o2_type": {
        "stage_id": "T+60",
        "type_label": "ransomware_like | non_ransom_coordinated | uncertain",
        "key_signals": [
        {"signal": "remote_service_exec", "host": "HOST-C", "evidence_id": "EVID-301"},
        {"signal": "multi_host_auth_burst", "hosts": ["HOST-A","HOST-B"], 
        "evidence_id": "EVID-210}
    ],
    "rationale": "1-3 sentences; must cite at least 2 distinct evidence objects"
    }
}
\end{verbatim}

Evidence examples that commonly support ransomware-like classification (must be grounded in the data sources):
\begin{itemize}[leftmargin=1.25em]
  \item Remote service execution / PsExec-like behavior (service-based remote execution patterns)~\cite{mitre_t1569_002}.
  \item SMB / admin-share style lateral movement artifacts~\cite{mitre_t1021_002}.
  \item Service creation artifacts (e.g., Event ID 7045) aligned with admin-tool execution traces~\cite{event_7045}.
\end{itemize}

\paragraph{O3. Cross-stage Incident Alert Triage Bundle (SOC output)}
\begin{itemize}[leftmargin=1.25em]
  \item \textbf{Goal:} Produce one SOC-style alert bundle that correlates cross-host and cross-stage signals and explains why this is \\
  (i) campaign-scale and \\
  (ii) ransomware-like (or not), using evidence.
  \item \textbf{Submit Part 1 (JSON):} list of correlated signals (no duplication of O1/O2 labels; focus on signal correlation).
\end{itemize}

\begin{verbatim}
{
   "o3_correlated_signals": [
    {
      "signal_type": "multi_host_anomaly",
      "hosts_involved": ["HOST-A", "HOST-B"],
      "evidence_id": "EVID-219",
      "note": "Authentication bursts across multiple hosts"
    },
    {
      "signal_type": "admin_tool_remote_exec",
      "host": "HOST-C",
      "evidence_id": "EVID-301",
      "note": "Unexpected remote service execution consistent with PsExec-like tooling"
    }
  ]
}
\end{verbatim}

\begin{itemize}[leftmargin=1.25em]
  \item \textbf{Submit Part 2 (one paragraph):} a single paragraph that includes:
  \begin{enumerate}
    \item earliest stage where activity becomes campaign-scale (cross-host justification required),
    \item earliest stage where signals become ransomware-like (type justification required),
    \item a short evidence-backed link between the two (why these together indicate coordinated threat activity).
  \end{enumerate}
\end{itemize}

\subsection{Blue Team Activity Dependency}\label{sec:fox-diagram}
%\textit{(Principles: DP1, DP2, DP3)}

Figure \ref{foxordering} shows the observed ordering and dependencies between real-world
blue team activities when fulfilling task FOX. 

\ReaderDisclaimer{Section \ref{sec:fox-diagram}}{Following design principle DP2, 
Figure \ref{foxordering} should not be used to prompt any agents.}

\begin{figure}[H]
\centering
\resizebox{\textwidth}{!}{%
\begin{tikzpicture}[
    >=Stealth,
    seq/.style={-Stealth, thick},
    dataflow/.style={-Stealth, thick, dashed},
    xref/.style={-Stealth, thick, dotted},
    box/.style={rounded corners, draw, thick, align=left,
      inner sep=4pt, fill=gray!3, minimum height=12mm, minimum width=38mm, text width=38mm}
]

% Row 1: Main workflow
\node[box] (ingest)  at (0, 0)    {Ingest SOC\\telemetry streams\\(host, network, SIEM)};
\node[box] (correlate) at (5.6, 0)  {Cross-host\\correlation\\analysis};
\node[box] (classify) at (11.2, 0)  {Campaign-scale\\\& type\\classification};

% Row 2: Supporting processes
\node[box] (evidence)    at (0, -3.5)   {Evidence\\tracking\\(chain-of-custody)};
\node[box] (scoring) at (5.6,-3.5) {Stage-level\\assessment\\(O1, O2)};
\node[box] (bundle)   at (11.2,-3.5){Alert triage\\bundle assembly\\(O3)};

% Main sequence flow
\draw[seq] (ingest) -- (correlate);
\draw[seq] (correlate) -- (classify);
\draw[seq] (classify) -- (bundle);

% Data flows
\draw[dataflow] (ingest) -- (evidence);
\draw[dataflow] (correlate) -- (scoring);
\draw[dataflow] (classify) -- (scoring);

% Cross-references
\draw[xref] (evidence) -- (scoring);
\draw[xref] (scoring) -- (bundle);

% Legend
\node[anchor=north west] (L) at (-0.2, -5.8) {\small \textbf{Legend:}};
\centering
\draw[seq] (-0.2, -6.3) -- (1.8, -6.3); \node[anchor=west] at (2.0, -6.3) {\small Sequence flow};
\draw[dataflow] (-0.2, -6.9) -- (1.8, -6.9); \node[anchor=west] at (2.0, -6.9) {\small Data dependency};
\draw[xref] (-0.2, -7.5) -- (1.8, -7.5); \node[anchor=west] at (2.0, -7.5) {\small Evidence linkage};
%\node[anchor=west] at (-0.2, -8.1) {\small \textbf Diagram shows observed order/dependencies for Task Fox evaluation (no method implied).};
\end{tikzpicture}%
}
\caption{Observed blue team activity dependency for Task Fox.}
\label{foxordering}
\end{figure}

\subsection{Scoreboard}\label{sec:fox-scoreboard}
%\textit{(Principles: DP3, DP4)}

\paragraph{Scoring granularity.} Task FOX is scored per stage (e.g., every 30 minutes). Each stage yields points for O1-O3 using the same ring model. Total score is the sum across all evaluated stages (100 points in total).

\subsubsection{Ring model (applies to O1, O2, O3)}\label{sec:fox-rings}
For each outcome at a stage, grade the best matching submission for that outcome:

\begin{table}[H]
\centering
\renewcommand{\arraystretch}{1.3}
\begin{tabularx}{0.85\linewidth}{l|c|X}
\toprule
\textbf{Ring} & \textbf{Pts} & \textbf{Meaning} \\
\midrule
Bullseye & 3 & Correct conclusion + evidence-backed justification \\
& & ($\geq$ 2 concrete evidence objects where applicable). \\[0.2cm]
Inner & 2 & Correct conclusion but missing one key justification element \\
& & (weak correlation, thin evidence, or vague tie). \\[0.2cm]
Outer & 1 & Directionally plausible but underspecified \\
& & (e.g., ``multi-host activity increasing'' without anchors). \\[0.2cm]
Miss & 0 & Incorrect conclusion or unjustified claim. \\
\bottomrule
\end{tabularx}

\vspace{0.4cm}
\caption{Ring-based scoring model for Task Fox outcomes.}
\end{table}

\subsubsection{O1: Scale Detection (39 pts in total)}\label{sec:fox-o1}
\begin{itemize}[leftmargin=1.25em]
  \item \textbf{What is graded:} stage-level scale\_label correctness plus evidence linkage.
  \item \textbf{Points:} across all stages, O1 contributes 39 points in total. Maximally 3 points can be earned during one stage.
  \item \textbf{Ring expectations:}
  \begin{itemize}
    \item Bullseye: correct label and impacted-host list consistent with telemetry; includes evidence objects that show cross-host correlation.
    \item Inner: correct label but host list incomplete or correlation not tightly justified.
    \item Outer: roughly correct direction (e.g., identifies multi-host trend) but no reliable anchors.
  \end{itemize}
\end{itemize}

\subsubsection{O2: Type Reasoning (39 pts in total)}\label{sec:fox-o2}
\begin{itemize}[leftmargin=1.25em]
  \item \textbf{What is graded:} type\_label correctness conditional on O1 being campaign-scale, plus quality of evidence-backed reasoning.
  \item \textbf{Points:} across all stages, O2 contributes 39 points in total. Maximally 3 points can be earned during one stage.
  \item \textbf{Ring expectations:}
  \begin{itemize}
    \item Bullseye: correct type call with at least two distinct supporting signals (e.g., remote service execution~\cite{mitre_t1569_002}, service install artifacts~\cite{event_7045}, SMB lateral movement context~\cite{mitre_t1021_002}).
    \item Inner: correct type call but reasoning relies on one weak signal or repeats O1 language without adding type-specific justification.
    \item Outer: hedged but directionally aligned (e.g., ``possible ransomware staging'') with thin support.
  \end{itemize}
\end{itemize}

\subsubsection{O3: SOC Alert Bundle (22 pts in total)}\label{sec:fox-o3}
\begin{itemize}[leftmargin=1.25em]
  \item \textbf{What is graded:} whether the alert bundle provides correlated signals and a coherent one-paragraph SOC summary that identifies: \\ (i) first campaign-scale stage, \\ (ii) first ransomware-like stage, and \\ (iii) cross-host and cross-stage justification.
  \item \textbf{Points:} across all stages, O3 contributes 22 points in total. 
  \item \textbf{Ring expectations:}
  \begin{itemize}
    \item Bullseye: correlated-signal JSON is non-redundant and evidence-grounded; paragraph correctly anchors both ``firsts'' and ties them together.
    \item Inner: mostly correct, but missing one of the three required paragraph components or correlation is weak.
    \item Outer: SOC-like text but not anchored (reads like generic incident prose).
  \end{itemize}
\end{itemize}

% \subsection{References}\label{sec:fox-refs}
% \begin{enumerate}[leftmargin=1.25em]
%   \item MITRE ATT\&CK. MITRE ATT\&CK: Service Execution (T1569.002). 2025. \url{https://attack.mitre.org/techniques/T1569/002/}. Accessed 2026-02-09.
%   \item MITRE ATT\&CK. MITRE ATT\&CK: SMB/Windows Admin Shares (T1021.002). 2025. \url{https://attack.mitre.org/techniques/T1021/002/}. Accessed 2026-02-09.
%   \item Ultimate Windows Security. Security Event 4625: An account failed to log on. 2025. \url{https://www.ultimatewindowssecurity.com/securitylog/encyclopedia/event.aspx?eventid=4625}. \\ Accessed 2026-02-09.
%   \item Invictus Incident Response. System Event 7045: A service was installed in the system. 2025. \url{https://www.invictus-ir.com/news/a-service-was-installed-in-the-system-7045}. Accessed 2026-02-09.
% \end{enumerate}
% Task Fox Ends

% ---------- SECTION 5: Task Goat (drop-in) ----------
\section{Task Goat}\label{sec:goat}

\subsection{Task Overview}\label{sec:overview}

%\textit{(Principles: SOC-first, Outcome-only, Durability.)}

\paragraph{Objective.} Evaluate a multi-agent AI system on \emph{file-system forensics} during a Colonial Pipeline–style enterprise ransomware incident in the business IT environment (Windows). The task checks the ability to: (i) label encryption at file and directory levels, (ii) roll up host/share impact, (iii) detect VSS tamper, (iv) attribute primary encryptor processes when telemetry exists, and (v) produce a one-page summary of main findings.

\paragraph{What must be submitted.} The AI system must produce \textbf{five outcomes} (details in Section~\ref{sec:outcomes}):
\begin{enumerate}[leftmargin=1.25em]
  \item O1: Encryption-state labels at file and directory levels.
  \item O2: Host/share impact aggregations.
  \item O3: VSS tamper facts (host, event, timing) with evidence.
  \item O4: Attribution of primary encryptor process trees (when available).
  \item O5: One-page \emph{summary of main findings} referencing O1--O4 claims (\textit{no new evidence in O5}).
\end{enumerate}

\paragraph{Where inputs come from.} Existing SOC data sources: file-system metadata and change journals (Sec.~\ref{sec:data-fs}), process execution telemetry (Sec.~\ref{sec:data-proc}), backup/VSS logs (Sec.~\ref{sec:data-vss}), optional SIEM alert stream (Sec.~\ref{sec:data-siem}), grader-facing reference file pairs (Sec.~\ref{sec:data-refpairs}), and late help-desk reports written in employee phrasing (Sec.~\ref{sec:data-helpdesk}). Claims must reference resolvable \texttt{evidence\_id}s in the pack’s chain-of-custody.

\textit{Reader-only note.} Following DP3, this task specifies \emph{outcomes only}. Methods are deliberately unspecified;  evaluation is by matching outcomes to ground truth via the scoreboard (Sec.~\ref{sec:scoreboard}).

% \subsubsection*{5.1a.\;Design Principles (benchmarking)}
% \begin{itemize}[leftmargin=1.25em]
%  \item \textbf{SOC-first ordering:} The unit reflects what a SOC % \emph{observes and reports}, not the attacker’s internal sequence.
%  \item \textbf{Outcome-only:} Submissions are judged by claims/artifacts against ground truth; no subtasks or logic are mandated.
%  \item \textbf{Hidden cross-task hints:} Inter-animal relations (GOAT, PANDA, FOX, TIGER, MOUSE) are \emph{not} disclosed in task text to encourage emergent coordination.
%  \item \textbf{Intentional incompleteness:} Some common signals (e.g., signature IDS rule-hits) may be withheld to avoid shortcutting via static signatures alone.
%  \item \textbf{Durability:} Specification relies on stable OS/forensic constructs so it remains valid without changes for many years.
%\end{itemize}

\begin{comment}

\subsection{Terminology}\label{sec:terms}
%\textit{(Principles: Durability; reused, plain terminology.)}
\begin{itemize}[leftmargin=1.25em]˜
  \item \textbf{Path:} absolute file or directory path (NTFS or SMB share).
  \item \textbf{Logical share:} Windows drive letter (e.g., \texttt{C:}) or UNC share (e.g., \verb|\\server\share|).
  \item \textbf{VSS snapshot:} Windows Volume Shadow Copy snapshot.
  \item \textbf{Encrypted file:} content transformed by ransomware.
  \item \textbf{Ransom note:} small textual file deployed by the actor indicating extortion terms.
\end{itemize} 

\end{comment} 

\subsection{Scope and Assumptions}\label{sec:scope}
%\textit{(Principles: SOC-first, Durability.)}
\begin{itemize}[leftmargin=1.25em]
  \item The incident dynamics mirror the \textbf{Colonial Pipeline} case (2021) \cite{ColonialWiki, CISA-AA21-131A, DOJ-DarkSide}. 
  \item IT network only; OT/ICS encryption is out of the scope. 
  \item Windows endpoints and file servers are dominant; NTFS and SMB shares are prevalent.
  \item Input is existing data sources; the task evaluates outputs at designer-defined \emph{checkpoints/stages} using the scoreboard (Section~\ref{sec:scoreboard}).  %not runtime performance.
\end{itemize}

%---------------
\subsection{Data Sources and Signals}\label{sec:data}
%\textit{(Principles: SOC-first, Intentional incompleteness; reader-only signal notes are not for agents.)}

\ReaderDisclaimer{Subsection \ref{sec:data}}{Following DP2, no extracted signals should be exposed to the AI system or included in prompts. 
%Agents infer from raw data; graders/pack designers may consult these when constructing ground truth.
}

\subsubsection{File-system metadata and change journals}\label{sec:data-fs}
\begin{itemize}[leftmargin=1.25em]
  \item \textbf{Metadata:} path, size, modified time, owner, extension; directory membership.
  \item \textbf{Change journal / audit:} create, rename, write events; rename cascades; per-process writer where available.
  \item \textbf{Signals (reader-only):} rapid per-process write bursts; mass renames; rare-extension spikes; directory-level dispersion of small notes; content entropy/compressibility rise after rewrites (when chunked content is available). See \cite{LuoTDSC} for encryption-loop characteristics; see \cite{ColonialWiki,CISA-AA21-131A} for ransom-note dispersion and widespread content impact.
\end{itemize}

\subsubsection{Process execution telemetry (EDR/XDR)}\label{sec:data-proc}
%\textit{(Principles: SOC-first, Outcome-only.) Suspicious applications trigger file-system operations; this subsection follows those processes.}
\begin{itemize}[leftmargin=1.25em]
  \item \textbf{Process tree:} parent-child, command line, signer, image hash.
  \item \textbf{Write-event summaries:} per-process counts and magnitudes of write operations over rolling windows; selected module load context.
  \item \textbf{Signals (reader-only):} writer attribution to suspected lineage; outlier write profiles aligned with encryption loops \cite{LuoTDSC}. 
\end{itemize} 
%\textit{Note on 4.1 vs 4.2.} 4.1 records OS/file-system events per path; 4.2 records \emph{process-centric} execution context and aggregated write activity. They are complementary.

%\ReaderDisclaimer{Subsection 4.2}{Signals are for readers only and must not be loaded into agents.}

\subsubsection{Backup/VSS logs}\label{sec:data-vss}
\begin{itemize}[leftmargin=1.25em]
  \item \textbf{VSS ops:} create/delete/disable events; backup service warnings/errors.
  \item \textbf{Signals (reader-only):} snapshot deletion/disable prior to mass encryption; restoration viability degradation \cite{CISA-AA21-131A,ColonialWiki}.
\end{itemize} 

%\ReaderDisclaimer{Subsection 4.3}{Presence/ordering of VSS operations is part of ground-truth construction and tolerances. Do not reveal specifics to agents.}

\subsubsection{SIEM alert stream}\label{sec:data-siem}

This data source holds the stream of SIEM alerts. 

%\begin{itemize}[leftmargin=1.25em]
%  \item \textbf{Alerts:} high-level summaries derived from lower-level sources; corroboration only.
%\end{itemize} 

\subsubsection{Reference file pairs (grader-only resource)}\label{sec:data-refpairs}
%\textit{(Principles: Outcome-only, Durability.)}
\begin{itemize}[leftmargin=1.25em]
  \item \textbf{Content:} representative pre-incident cleartext files and matching post-incident encrypted counterparts included in the replay pack.
  \item \textbf{Checks (reader-only):} hash deltas; header/magic changes; entropy/compressibility shifts; directory-level encrypted-byte fractions aligned with the file/path inventory in Sec.~\ref{sec:data-fs}. See \cite{LuoTDSC}. 
  %\item \textbf{Exposure:} Available to graders and pack designers only; never surfaced to participating agents.
\end{itemize}

%\ReaderDisclaimer{Subsection 4.5}{Exact file selections, hashes, and byte-level comparators establish ground truth and thresholds. Do not reveal filenames or checksums to agents.}

\subsubsection{Helpdesk incident reports (employee phrasing)}\label{sec:data-helpdesk}
%\textit{(Principles: SOC-first; plain, non-technical descriptions.)}
\begin{itemize}
  \item \textbf{Content:} tickets or messages from employees describing symptoms such as ``files will not open,'' ``text looks scrambled,'' or ``ransom note on desktop.''
  \item \textbf{Signals (reader-only):} human-reported and typically later than telemetry; solid for narrative context and scoping. %not for exact timing or byte-level state.
  %\item \textbf{Scope:} not part of ground-truth computation for encryption labels; phrasing examples are illustrative and must not coach agents.
\end{itemize} 

%\ReaderDisclaimer{Subsection 4.6}{Do not call these reports ``artifacts,'' and do not feed example text to agents.}

%---------------
\subsection{Outcomes}\label{sec:outcomes}
%\textit{(Principles: Outcome-only; no agent advice.)}

\textbf{O1.\;Encryption-state labels.}
\emph{Truth.} A file is \emph{encrypted} if its bytes match the post-encryption reference (Sec.~\ref{sec:data-refpairs}) and/or a decryptor/restore is required; a directory is \emph{encrypted} if its encrypted-byte fraction exceeds the manifest threshold.\\
\emph{Submit.} JSON list \texttt{encrypted\_paths} with \{\texttt{path}, \texttt{label}\(\in\)\{encrypted, not-yet-encrypted, unknown\}, \texttt{evidence\_id}\}.

\textbf{O2.\;Host/share impact.}
\emph{Truth.} Fraction of bytes encrypted and an impacted boolean per share/host derived from O1.\\
\emph{Submit.} JSON \texttt{impact\_by\_host\_share}: \{\texttt{host}, \texttt{share}, \texttt{encrypted\_bytes}, \texttt{total\_bytes}, \texttt{first\_seen}, \texttt{encrypted\_bytes\_fraction}\}.

\textbf{O3.\;VSS tamper.}
\emph{Truth.} Snapshot delete/disable events prior to or during encryption waves on affected hosts.\\
\emph{Submit.} JSON \texttt{vss\_events}: \{\texttt{host}, \texttt{event\_type}, \texttt{time}, \texttt{evidence\_id}\}.

\textbf{O4.\;Attribution (when available).}
\emph{Truth.} Primary encryptor lineage is the process tree responsible for the majority of encrypted bytes per host.\\
\emph{Submit.} JSON \texttt{attribution}: \{\texttt{host}, \texttt{pid}, \texttt{parent\_pid}, \texttt{cmdline}, \texttt{bytes\_written}, \texttt{evidence\_id}\}.

\textbf{O5.\;One-page summary of main findings.}
\emph{Truth.} A single-page summary accurately reflects O1–O4 and introduces no new claims or evidence. It may cite \texttt{evidence\_id}s that already appear in the O1–O4 JSON.\\
\emph{Submit.} One-page PDF summarizing top impacts/findings and referencing O1–O4 (by \texttt{evidence\_id} or claim identifiers); do not include raw evidence in O5.

%----------
\subsection{Blue Team Activity Dependency}\label{sec:diagram}

\begin{figure}[H]
\centering
\resizebox{\textwidth}{!}{%
\begin{tikzpicture}[
    >=Stealth,
    seq/.style={-Stealth, thick},
    dataflow/.style={-Stealth, thick, dashed},
    xref/.style={-Stealth, thick, dotted},
    box/.style={rounded corners, draw, thick, align=left,
      inner sep=4pt, fill=gray!3, minimum height=12mm, minimum width=38mm, text width=38mm}
]
\node[box] (ingest)  at (0, 0)    {Ingest data packs\\(FS, EDR, VSS, SIEM, help-desk)};
\node[box] (outcomes) at (5.6, 0)  {Produce outcomes\\(O1--O4 data products)};
\node[box] (summary) at (11.2, 0)  {Assemble O5\\(one-page findings)};

\node[box] (qc)    at (0, -3.5)   {Quality checks\\(evidence links, CoC)};
\node[box] (rollup) at (5.6,-3.5) {Aggregations\\(host/share, incidents)};
\node[box] (emit)   at (11.2,-3.5){Emit checkpoint\\submissions};

\draw[seq] (ingest) -- (outcomes);
\draw[seq] (outcomes) -- (summary);
\draw[seq] (summary) -- (emit);

\draw[dataflow] (ingest) -- (qc);
\draw[dataflow] (outcomes) -- (rollup);
\draw[dataflow] (rollup) -- (summary);

\draw[xref] (qc) -- (summary);

\node[anchor=north west] (L) at (-0.2, -5.6) {\small \textbf{Legend:}};
\draw[seq] (-0.2, -6.1) -- (1.8, -6.1); \node[anchor=west] at (2.0, -6.1) {\small Sequence};
\draw[dataflow] (-0.2, -6.7) -- (1.8, -6.7); \node[anchor=west] at (2.0, -6.7) {\small Data feed};
\draw[xref] (-0.2, -7.3) -- (1.8, -7.3); \node[anchor=west] at (2.0, -7.3) {\small Cross-ref to evidence};
%\node[anchor=west] at (-0.2, -7.9) {\scriptsize Diagram shows observed order/dependencies only (no method implied).};
\end{tikzpicture}%
}
\caption{Observed blue team activity dependency for Task Goat.}
\label{fig:goatorder}
\end{figure}
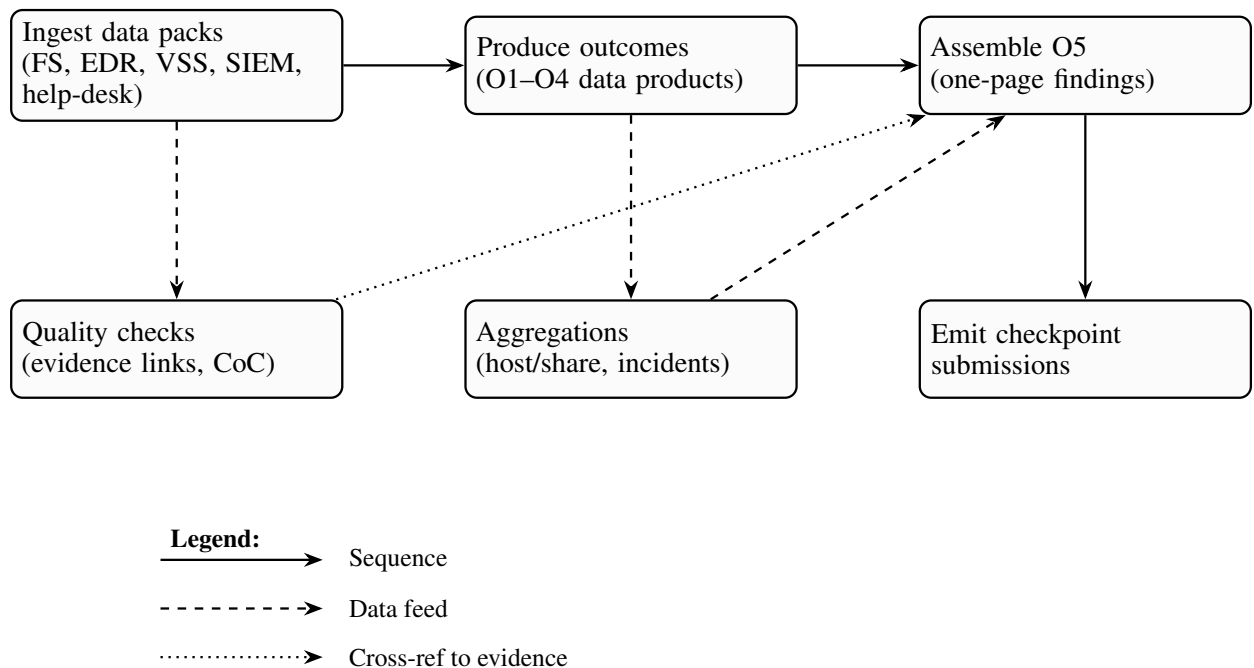

Figure \ref{fig:goatorder} shows the observed ordering and dependencies between real-world
blue team activities when fulfilling task Goat. 

\ReaderDisclaimer{Section \ref{sec:diagram}}{Following design principle DP2, 
Figure \ref{fig:goatorder} should not be used to prompt any agents.}

%\textit{(Principles: SOC-first; observed order and dependencies only, no method implied.)}
%\ReaderDisclaimer{Section 6}{Diagram reflects \emph{observed} ordering and dependencies in replay packs. It does not prescribe any method and should not be used to coach agents.}

\subsection{Scoreboard}\label{sec:scoreboard}
%\textit{(Principles: SOC-first, Outcome-only.)}

At each checkpoint, the manifest defines \emph{targets} for O1–O5. Submissions provide \emph{claims} with \texttt{evidence\_id}s resolvable to pack sources. Scoring is ring-based per target; penalties are applied within the same flow so graders do not “finish then revisit.” All numeric/time matching follows manifest tolerances known to graders (not to agents).

\subsubsection{Rings and matching (applies to O1--O4)}\label{sec:rings}
For each target, consider only the best matching claim:
\begin{itemize}[leftmargin=1.25em]
  \item \textbf{File-level exact:} correct file with required detail. \hfill  \(\textit{Exact}\)
  \item \textbf{Directory-level:} correct directory but not the specific file, or minor incompleteness at directory granularity. \hfill  \(\textit{Directory}\)
  \item \textbf{Host/share-level:} only the correct host/share is indicated. \hfill  \(\textit{Host/Share}\)
  \item \textbf{Miss:} incorrect or unsupported. \hfill  \(\textit{Miss}\)
\end{itemize}

\textbf{Mistakes to avoid (grader).}
Do not map multiple claims to the same target; do not award any ring unless a valid \texttt{evidence\_id} resolves in chain-of-custody; do not infer correctness from SIEM summaries alone.

\subsubsection{General penalties (apply as you grade)}\label{sec:penalties}
\begin{itemize}[leftmargin=1.25em]
  \item \textbf{Wrong assertions:} each wrong claim that asserts a state (e.g., marks a path as encrypted when manifest disagrees) incurs \(-1\) for that outcome, capped at 20\% of the outcome’s max at that checkpoint.
  \item \textbf{No evidence:} claims lacking a resolvable \texttt{evidence\_id} are ineligible for ring points and incur \(-1\) each (cap 10\% of outcome max).
  \item \textbf{Contradictions:} mutually contradictory claims within the same stage about the same target yield \(-2\) for that pair.
  \item \textbf{Stage leakage:} citing evidence that appears only at a later stage to justify a current-stage claim yields \(-2\) per occurrence.
  \item \textbf{Duplicate spam:} duplicated claims beyond the first for the same target incur \(-0.5\) each (cap 10\% of outcome max).
  \item \textbf{Over-submission spam:} let \(B_X\) be the budget (max evaluated claims) for outcome \(X\) at a stage, and let \(C_X\) be total submitted claims for \(X\). Define a spam threshold \(S_X = 2B_X\). If \(C_X > S_X\), apply
  \[
     P_{\text{spam}}(X) \;=\; -\,\Big\lfloor \frac{C_X - S_X}{B_X} \Big\rfloor \,,
  \]
  i.e., \(-1\) per full extra budget beyond \(2\times\) over-submission.
\end{itemize}

\subsubsection{Outcome-specific targets, points, and common grading mistakes}\label{sec:outcome-points}

\paragraph{O1\;Encryption-state labels (40 pts).}
Targets: 20 file/dir items. Ring mapping per item:
\(\textit{Exact}=2.0\), \(\textit{Directory}=1.0\), \(\textit{Host/Share}=0.5\), \(\textit{Miss}=0\).\\
\emph{Common mistakes (penalize as in Sec.~\ref{sec:penalties}):} trusting ransom-note presence alone; labeling a directory encrypted without meeting the manifest threshold; claiming \emph{unknown} when truth is known.

\paragraph{O2\;Host/share impact (25 pts).}
Targets: 5 impact facts with encrypted-byte fraction intervals. Per fact:
\(\textit{Exact}=5\) (correct scope and interval overlap), \(\textit{Directory}=3\) (scope correct, interval off), \(\textit{Host/Share}=1\), \(\textit{Miss}=0\).\\
\emph{Common mistakes:} aggregating across hosts; ignoring \texttt{first\_seen}; intervals that do not overlap the manifest interval.

\paragraph{O3\;VSS tamper (15 pts).}
Targets: 3 host-level facts \((\text{host}, \text{event}, \text{time})\). Per fact:
\(\textit{Exact}=5\), \(\textit{Directory}=3\) (host correct, timing vaguely stated), \(\textit{Host/Share}=1\) (only host impacted), \(\textit{Miss}=0\).\\
\emph{Common mistakes:} conflating generic backup errors with VSS snapshot delete/disable; timing after encryption waves.

\paragraph{O4\;Attribution (10 pts).}
Targets: 2 host-level attribution facts (top writer process tree with parent/cmdline). Per fact:
\(\textit{Exact}=5\), \(\textit{Directory}=3\) (lineage partially specified), \(\textit{Host/Share}=1\), \(\textit{Miss}=0\).\\
\emph{Common mistakes:} attributing to a helper tool instead of the encryptor lineage; omitting parent or cmdline; relying on SIEM text without process evidence.

\paragraph{O5\;One-page summary (10 pts).}
Checklist items (binary, 2 pts each): (i) covers O1 labels, (ii) covers O2 impact with numbers, (iii) covers O3 with host/time, (iv) covers O4 when telemetry exists, (v) references O1–O4 by claim/evidence identifiers without introducing new evidence in O5.\\
\emph{Common mistakes:} marketing language with no references to submitted claims; numbers inconsistent with O1–O4 JSON; placing raw evidence or new claims in O5.

\subsection{Compliance and Safety}\label{sec:compliance}
%\textit{(Principles: Durability, Outcome-only.)}
Read-only analysis; no actions against hosts. All claims must reference \texttt{evidence\_id} entries resolvable in the pack. Reader-only subsections are for graders and must not coach participating systems.

% ---------- end Section 5 ----------

\section{Task Mouse}
\subsection{Task Overview}
\paragraph{Objective.}
%In this task, you are required to design and implement 
Evaluate a multi-agent AI system's ability to 
analyze heterogeneous system and network telemetry to determine whether a data exfiltration event has occurred in the detected large-scale attack campaign.  
The AI system is required to produce five outcomes and will be evaluated according to a scoreboard.
\paragraph{Input.} Following DP1 and DP4, the heterogeneous system and network telemetry (detailed in Section \ref{mouse:data}) includes: packet capture (PCAP) files, firewall logs, Windows event logs, and Linux journals.
\paragraph{Required Outcomes.} Following DP3 and DP5, the five outcomes that the AI agent must produce are (detailed in \ref{mouse:scoreboard}):
\begin{itemize}
    \item O1: Determination of Exfiltration Occurrence (JSON file)
    \item O2: Estimation of Exfiltration Start Time (JSON file)
    \item O3: Estimation of Exfiltrated Data Volume (JSON file)
    \item O4: Identification of Hosts Involved in Exfiltration (JSON file)
    \item O5: Identification of Exfiltration Protocols (JSON file)
\end{itemize}

\subsection{Scoreboard Design}\label{mouse:scoreboard}
The AI system must produce five outcomes. Each outcome is evaluated independently according to the success and failure definitions below. If \texttt{"exfiltration\_occurred"} is \texttt{"No"}, all remaining outcomes must be set to \texttt{"N/A"} and only the first outcome is scored.

%%%%%%%%%%%%%%%%%%%%%%%%%%%%%%%%%%%%%%%%%%%%%%%%%%%%%%%%%%%%%%%%%%%%%%%%%%%%
% OUTCOME 1
%%%%%%%%%%%%%%%%%%%%%%%%%%%%%%%%%%%%%%%%%%%%%%%%%%%%%%%%%%%%%%%%%%%%%%%%%%%%

\subsubsection{O1: Determination of Exfiltration Occurrence (20 points)}%
\begin{description}
    \item[Purpose] Determine whether a data-exfiltration event occurred in the attack campaign.
    \item[Allowed values] \texttt{"Yes"} or \texttt{"No"}.
    \item[Success (20 pts)] The value exactly matches the ground truth. 
    \item[Failure (0 pts)] The value does not match the ground truth or is left blank.
    \item[Required outcome format]
\end{description} 
\begin{lstlisting}
// Always required
{
  "exfiltration_occurred": "Yes"
}
\end{lstlisting}

%%%%%%%%%%%%%%%%%%%%%%%%%%%%%%%%%%%%%%%%%%%%%%%%%%%%%%%%%%%%%%%%%%%%%%%%%%%%
% OUTCOME 2
%%%%%%%%%%%%%%%%%%%%%%%%%%%%%%%%%%%%%%%%%%%%%%%%%%%%%%%%%%%%%%%%%%%%%%%%%%%%

\subsubsection{O2: Estimation of Exfiltration Start Time (20 points)}%
\begin{description}
    \item[Purpose] Identify when exfiltration began, i.e., when data first left the compromised environment.
    \item[Allowed values] ISO-8601 timestamp at minute precision: YYYY-MM-DDTHH:MMZ.
    \item[Success (20 pts)] Timestamp is within $\pm 5$ minutes of the ground truth.
    \item[Failure (0 pts)] Deviation greater than 5 minutes, malformed timestamp, or blank entry.
    \item[Required outcome format]
\end{description}
\begin{lstlisting}
// If exfiltration_occurred == "Yes"
{
  "estimated_start_time": "2025-03-21T14:37Z"
}

// If exfiltration_occurred == "No"
{
  "estimated_start_time": "N/A"
}
\end{lstlisting}

%%%%%%%%%%%%%%%%%%%%%%%%%%%%%%%%%%%%%%%%%%%%%%%%%%%%%%%%%%%%%%%%%%%%%%%%%%%%
% OUTCOME 3
%%%%%%%%%%%%%%%%%%%%%%%%%%%%%%%%%%%%%%%%%%%%%%%%%%%%%%%%%%%%%%%%%%%%%%%%%%%%

\subsubsection{O3: Estimation of Exfiltrated Data Volume (20 points)}%
\begin{description}
    \item[Purpose] Approximate the total volume of data exfiltrated.
    \item[Allowed values] Numeric gigabyte value, integer approximation preferred.
    \item[Success (20 pts)] Value is within $\pm 10$ GB of the ground truth.
    \item[Failure (0 pts)] Deviation greater than 10 GB, non-numeric, or blank entry.
    \item[Required outcome format]
\end{description}
\begin{lstlisting}
// If exfiltration_occurred == "Yes"
{
  "estimated_data_volume_gb": 95
}

// If exfiltration_occurred == "No"
{
  "estimated_data_volume_gb": "N/A"
}
\end{lstlisting}

%%%%%%%%%%%%%%%%%%%%%%%%%%%%%%%%%%%%%%%%%%%%%%%%%%%%%%%%%%%%%%%%%%%%%%%%%%%%
% OUTCOME 4
%%%%%%%%%%%%%%%%%%%%%%%%%%%%%%%%%%%%%%%%%%%%%%%%%%%%%%%%%%%%%%%%%%%%%%%%%%%%

\subsubsection{O4: Identification of Hosts Involved in Exfiltration (20 points)}%
\begin{description}
    \item[Purpose] Enumerate hosts directly involved in staging (collecting, compressing) or transmitting exfiltrated data.
    \item[Allowed values] JSON array of hostnames or IP addresses.
    \item[Success (20 pts)] Reported set overlaps with the ground truth, tolerating up to three additional or missing hosts.
    \item[Failure (0 pts)] Deviation greater than three hosts or malformed list.
    \item[Required outcome format]
\end{description}
\begin{lstlisting}
// If exfiltration_occurred == "Yes"
{
  "involved_hosts": [
    "host-12.corp.local",
    "10.2.3.45"
  ]
}

// If exfiltration_occurred == "No"
{
  "involved_hosts": "N/A"
}
\end{lstlisting}

%%%%%%%%%%%%%%%%%%%%%%%%%%%%%%%%%%%%%%%%%%%%%%%%%%%%%%%%%%%%%%%%%%%%%%%%%%%%
% OUTCOME 5
%%%%%%%%%%%%%%%%%%%%%%%%%%%%%%%%%%%%%%%%%%%%%%%%%%%%%%%%%%%%%%%%%%%%%%%%%%%%

\subsubsection{O5: Identification of Exfiltration Protocols (20 points)}%
\begin{description}
    \item[Purpose] List all application-layer or management protocols used to perform exfiltration.
    \item[Allowed values] Uppercase protocol identifiers in a JSON array; use \texttt{["UNKNOWN"]} if the protocol cannot be inferred.
    \item[Success (20 pts)]  
    All true exfiltration protocols are listed and no incorrect protocols appear.
    \item[Partial Success (10 pts)]  
    The primary (major) exfiltration protocol is correctly listed, but one or more additional protocols are incorrectly included or omitted.
    \item[Failure (0 pts)]  
    The primary (major) exfiltration protocol is missing, or the entry is blank, or the listed protocols are all incorrect.
    \item[Required outcome format]
\end{description}
\begin{lstlisting}
// If exfiltration_occurred == "Yes"
{
  "exfiltration_protocols": [
    "HTTPS",
    "DNS"
  ]
}

// If exfiltration_occurred == "No"
{
  "exfiltration_protocols": "N/A"
}
\end{lstlisting}

\subsection{Data Sources}\label{mouse:data}

The agent will receive as input a unified dataset drawn from the following system and network telemetry.

%The data sources encompass multiple layers of system and network telemetry. Network flows provide summaries of communication sessions, including duration, volume, and endpoints. Firewall logs capture connection attempts and rule matches at network boundaries. Host event logs record operating system–level activities such as authentication, process creation, and scheduled tasks. System logs include low-level records such as system calls and security events that offer additional visibility into host behaviors.

\subsubsection{PCAP Files}\label{sec:pcap}

The AI agent receives \emph{raw packet-capture files} in standard \texttt{pcap} or \texttt{pcapng} format. These captures contain network packets collected directly from a boundary router.

To illustrate the type of information contained in a decoded packet, we show below a sample produced by \texttt{tcpdump}~\cite{tcpdump} %\footnote{TCPdump is a command-line packet capture and network traffic inspection tool that records and prints raw packets directly from an interface. See https://github.com/the-tcpdump-group/tcpdump/.}, 
which parses a raw PCAP file into a text representation.

\begin{lstlisting}[language=bash,caption={Example packet decoded from a raw PCAP file using \texttt{tcpdump}.}]
20:15:43.512345 IP (tos 0x0, ttl 52, id 54321, offset 0, flags [DF], proto TCP (6))
10.0.2.15.44562 > 93.184.216.34.80: Flags [S], cksum 0x1c2d (correct),
seq 123456789, win 64240, options [mss 1460,sackOK,TS val 123456 ecr 0,nop,wscale 7], length 0
\end{lstlisting}

Note that the PCAP example is generated by a decoding tool and is not the raw capture. The AI agent receives the original \texttt{.pcap} file.

\subsubsection{Firewall Logs}\label{sec:firewall_logs}

The AI agent receives \emph{raw firewall logs} exported directly from network firewall appliances deployed at the enterprise perimeter. These records are plain-text log entries generated by firewall appliances. They may be exported to monitoring systems using various mechanisms, but the AI agent receives the raw text log entries themselves.

Listing \ref{lst:firewall log} show an example of a single firewall log entry collected from a Palo Alto Networks PA-Series firewall~\cite{panos-syslog}. This representation is produced by Palo Alto's standard syslog export format and is shown here solely as an example of decoded log structure. The AI agent receives the original raw text logs rather than this rendered formatting.

\begin{lstlisting}[language=bash,caption={Example Palo Alto Networks firewall log entry exported via syslog.}]
Feb 13 11:42:31 PA-3050 1,2023/02/13 11:42:31,001801000012,TRAFFIC,start,2305, 2023/02/13 11:42:31,
10.1.5.23,8.8.8.8,0.0.0.0,0.0.0.0,Allow-DNS,,,dns,vsys1, ethernet1/3,ethernet1/4,
LogForwardingProfile,202.55.0.2,10.1.5.23,0,24576,53, udp,allow,0,0,0,0,,PA-3050,from-policy
\end{lstlisting}\label{lst:firewall log}

This entry corresponds to a DNS outbound session allowed by a PA-3050 firewall and includes key fields such as the firewall serial number, log type (\texttt{TRAFFIC}), session state (\texttt{start}), timestamps, source and destination IP addresses, ports, protocol (\texttt{udp}), security rule (\texttt{Allow-DNS}), ingress and egress interfaces, NAT translation fields, and the final action taken by the firewall.

\subsubsection{Windows Event Logs}\label{sec:windows_event_log}

The AI agent receives \emph{raw Windows Event Log files} in the native \texttt{.evtx} format~\cite{windows-event-log}. 
These files contain binary-encoded event records generated by the Windows Event Logging subsystem. The \texttt{.evtx} structure includes record headers, provider metadata, template definitions, localized message references, and binary payloads. None of these elements are directly human-readable in their raw form. 

To illustrate the type of information found in Windows host events, we show in Table \ref{tab:windows_event_fields} the parsed representation of a specific Sysmon~\cite{sysmon} Event~7 record. This example is produced  using the \texttt{Chainsaw} tool~\cite{chainsaw} %\footnote{Chainsaw is a tool for rapidly triaging Windows forensic artefacts like EVTX logs using Sigma or custom detection rules. See https://github.com/WithSecureLabs/chainsaw/.}, 
which extracts and renders the XML event structure into a  readable form.

\label{tab:windows_event_fields}
\begin{table*}[htbp]
\centering
\footnotesize
\setlength{\tabcolsep}{4pt}
\renewcommand{\arraystretch}{1.05}
\caption{Example fields extracted from a Sysmon Event~7 record using Chainsaw.}
\begin{tabular}{p{0.28\linewidth}p{0.68\linewidth}}
\toprule
\textbf{Field} & \textbf{Value / Interpretation} \\
\midrule

\texttt{EventID} & \texttt{7} (Sysmon ImageLoad event). \\

\texttt{Provider.Name} & \texttt{Microsoft-Windows-Sysmon}. \\

\texttt{Provider.Guid} & \texttt{5770385F-C22A-43E0-BF4C-06F5698FFBD9}. \\

\texttt{Version} & Event schema version (\texttt{3} in this instance). \\

\texttt{Level} & Severity or verbosity level (\texttt{4}). \\

\texttt{Task} & Subcategory associated with this event (\texttt{7}). \\

\texttt{Opcode} & Operational opcode (\texttt{0}). \\

\texttt{Keywords} & Bitmask classification (\texttt{0x8000000000000000}). \\

\texttt{TimeCreated.SystemTime} &
Timestamp of the event in UTC (\texttt{2020-10-13T20:11:42.269224Z}). \\

\texttt{EventRecordID} & Sequence number within the log (\texttt{2196441}). \\

\texttt{Execution.ProcessID} & Process ID of the event emitter (\texttt{5340}). \\

\texttt{Execution.ThreadID} & Thread ID of the event emitter (\texttt{7092}). \\

\texttt{Channel} &
Logical channel from which the event originates (\texttt{Microsoft-Windows-Sysmon/Operational}). \\

\texttt{Computer} & Hostname of the system emitting the event (\texttt{LAPTOP-JU4M3I0E}). \\

\texttt{Security.UserID} & Security Identifier of the actor (\texttt{S-1-5-18}). \\

\texttt{UtcTime} &
Timestamp provided within Sysmon event data (\texttt{2020-10-13 20:11:42.268}). \\

\texttt{ProcessGuid} & Sysmon GUID for process instance. \\

\texttt{ProcessId} & ID of the process that loaded the image (\texttt{1716}). \\

\texttt{Image} & Executable responsible for the load (\texttt{C:\textbackslash Windows\textbackslash System32\textbackslash wuauclt.exe}). \\

\texttt{ImageLoaded} & DLL or module loaded (\texttt{C:\textbackslash ProgramData\textbackslash Intel\textbackslash helpa.dll}). \\

\texttt{FileVersion / Description / Product / Company / OriginalFileName} &
Metadata extracted from the loaded file; may be empty or unavailable. \\

\texttt{Hashes} & Cryptographic hashes (SHA1, MD5, SHA256, IMPHASH) computed by Sysmon. \\

\texttt{Signed} & Indicates whether the loaded image is digitally signed. \\

\texttt{Signature / SignatureStatus} &
Signature information when available. \\
\bottomrule
\end{tabular}
\end{table*}

Note that the Windows event log example is produced by a parser and is not the native \texttt{.evtx} content. The AI agent receives the original \texttt{.evtx} file.

\subsubsection{Linux Journals}\label{sec:linux_journald}

The AI agent receives \emph{raw systemd--journald log files (journals)} in their native binary
\texttt{.journal} format~\cite{journald}. These files record host-level events produced by the Linux kernel, system services managed by \texttt{systemd}, authentication components, user processes, and optional subsystems such as \texttt{auditd}. Although journald stores logs in a binary format, each record conceptually consists of a set of key–value fields defined by the emitting service or subsystem~\cite{journald-fields}. 
Because journald imposes no fixed schema, different events share only a small subset of common metadata.

Typical fields observed in journald records include:
\begin{itemize}
    \item \textbf{Core metadata:}\\
    {\ttfamily\small
        \_MACHINE\_ID, \_HOSTNAME, \_BOOT\_ID, \_\_REALTIME\_TIMESTAMP, 
        \_\_MONOTONIC\_TIMESTAMP, \_\_CURSOR, PRIORITY, MESSAGE
    }

    \item \textbf{Process-related fields:}\\
    {\ttfamily\small
        \_PID, \_UID, \_GID, \_CMDLINE, \_EXE, \_COMM, \_CAP\_EFFECTIVE
    }

    \item \textbf{Systemd service metadata:}\\
    {\ttfamily\small
        \_SYSTEMD\_UNIT, \_SYSTEMD\_SLICE, \_SYSTEMD\_CGROUP, 
        \_SYSTEMD\_INVOCATION\_ID
    }

    \item \textbf{Kernel and device fields:}\\
    {\ttfamily\small
        \_TRANSPORT=kernel, \_KERNEL\_DEVICE, \_KERNEL\_SUBSYSTEM,
        \_UDEV\_SYSNAME
    }

    \item \textbf{Audit fields (when \texttt{auditd} is enabled):}\\
    {\ttfamily\small
        \_AUDIT\_TYPE, \_AUDIT\_SESSION, \_AUDIT\_LOGINUID, \_AUDIT\_FIELD\_*
    }
\end{itemize}

The example in Listing \ref{lst:linux journal} illustrates two journald records produced by the Linux kernel during system boot. These entries were obtained by exporting a raw \texttt{.journal} file using:
\[
\texttt{journalctl --file /path/to/system.journal -o json-pretty}
\]

\begin{lstlisting}[caption={Example journal records parsed via \texttt{journalctl}.}]
{
    "__MONOTONIC_TIMESTAMP": "7541720",
    "SYSLOG_IDENTIFIER": "kernel",
    "__SEQNUM": "1",
    "_BOOT_ID": "13bbbc3f31e94c228116fea9d3b72e12",
    "_SOURCE_MONOTONIC_TIMESTAMP": "0",
    "__CURSOR": "s=52fdc7e1551b42e8928d7230e83eeaef;..."
    "SYSLOG_FACILITY": "0",
    "_TRANSPORT": "kernel",
    "_MACHINE_ID": "a053b0b5b42548de8818448a7995d4b5",
    "__REALTIME_TIMESTAMP": "1612270832256460",
    "MESSAGE": "Linux version 5.4.0-65-generic (buildd@lcy01-amd64-018)...",
    "__SEQNUM_ID": "52fdc7e1551b42e8928d7230e83eeaef",
    "PRIORITY": "5",
    "_HOSTNAME": "ubuntu"
}
{
    "_MACHINE_ID": "a053b0b5b42548de8818448a7995d4b5",
    "SYSLOG_IDENTIFIER": "kernel",
    "PRIORITY": "6",
    "__MONOTONIC_TIMESTAMP": "7541744",
    "__REALTIME_TIMESTAMP": "1612270832256485",
    "_TRANSPORT": "kernel",
    "_SOURCE_MONOTONIC_TIMESTAMP": "0",
    "_BOOT_ID": "13bbbc3f31e94c228116fea9d3b72e12",
    "__CURSOR": "s=52fdc7e1551b42e8928d7230e83eeaef;...",
    "MESSAGE": "Command line: BOOT_IMAGE=/boot/vmlinuz-5.4.0-65-generic ...",
    "__SEQNUM_ID": "52fdc7e1551b42e8928d7230e83eeaef",
    "SYSLOG_FACILITY": "0",
    "_HOSTNAME": "ubuntu",
    "__SEQNUM": "2"
}
\end{lstlisting}\label{lst:linux journal}

Note that the Linux journal example is the output of a journal-decoding tool and is not the raw \texttt{.journal} data. The AI agent receives the original \texttt{.journal} file.

\subsection{Remarks}
\begin{tightitemize}
    \item Following DP2, this task does not disclose information about other tasks in SOC-bench.
    \item Unlike Windows Event Log~\cite{windows-event-log}, Linux journald~\cite{journald-fields} does not enforce a stable event schema. Windows host events always include rich, structured fields (e.g., process IDs, user IDs, image names, status codes), whereas Linux provides only minimal default metadata. Comparable detail appears in Linux logs only when the \textsc{auditd} subsystem is enabled and appropriate audit rules are configured. 
\end{tightitemize}

\section{Task Tiger}
\subsection{Task Overview}
\textbf{Objective.} Evaluate an AI agent’s ability to analyze indicators of compromise (IoCs) to attribute the attack, then report any used Tactics, Techniques, and Procedures (TTPs). This task tests whether the AI can provide sufficient and actionable decision-making as a senior Tier 2 SOC analyst. 
\begin{itemize}
\item Real-world Tier 2 SOC analysts are responsible for accurate forensics and incident response after Tier 1 performs initial triage. Following \textit{DP1} of SOC-bench, Task Tiger will not penalize agents based on time to complete. As Tier 2 SOC analysts are in a more senior role, accurate decision-making is more valuable than minimum time.
\item Real-world SOCs still have time pressure and want to be fast. However, human Tier 2 analysts cannot achieve accuracy without sufficient time.
\item Agents will still receive feedback at the standard interval of every 30 minutes. 
\end{itemize}
\textbf{Expected Outcomes.} The agent must produce three deliverables:
\begin{enumerate}
\item O1: List of data sources that apply to the investigation and their relationships (JSON)
\begin{verbatim}
{
    "Windows Event Logs": {
        "Correlated Source": [ "SIEM Logs", "Firewall Logs" ] 
    },
    "CPU Temperature": {
        "Correlated Source": [ "Disk Images" ] 
    },    
}
\end{verbatim}
\item O2: Threat Graph data structure representing the progression of events with evidence, such as security flaws, from other data sources (JSON)
\begin{verbatim}
{
    "Node-<ID>": {
        "Artifact": "Mailserver <IP Address>",
        "Vulnerabilities": "Weak Administrator Password",
        "Data Source Attribution": [ "SIEM Logs", "Firewall Logs" ],
    },
    "Edge-<ID>": {
        "Parent": [ "Node-<ID>" ],
        "Child": [ "Node-<ID>" ],
        "Interaction": "Malicious process <Name.exe> encrypting files on <Mailserver>",
        "Data Source Attribution": [ "IDS Logs" ],
    },
    "Node-<ID>": {
        "Artifact": Executable <Malware.exe>"
        "Vulnerabilities": "",
        "Data Source Attribution": [ "SIEM Logs" ],
    },
}
\end{verbatim}
\item O3: One-paragraph statement that identifies the initial entrypoint for the compromise (Plaintext)

\end{enumerate}
\fbox{%
  \parbox{\textwidth}{%
    \textbf{Warning:} Presence of a data source in the following section does not guarantee that it is relevant to Task Tiger.
  }%
}
\subsection{Data Sources}
Task Tiger represents the final phase of incident response that is assigned to Tier 2 SOC analysts. Therefore, agents may apply findings from preceding tasks (Fox, Tiger, Goat, Mouse) to their reasoning. The available data sources are observed from the Colonial Pipeline incident, in accordance with \textit{DP3} of SOC-bench. Agents may reference the following data sources for decision-making and evidence:

\begin{enumerate}
\item Cyber Threat Intelligence (CTI) Feeds
    \begin{itemize}
        \item Identify relevant information related to the ransomware attack (Colonial Pipeline), such as password dumps, previous data breaches
        \item Trap: Hidden among dozens of unrelated CTI Sources
    \end{itemize}
\item Packet Capture (PCAP) Files
    \begin{itemize}
        \item Identify malicious actions from raw packet data, such as timestamps, abused protocols (HTTP, FTP), etc.
        \item Trap: Include thousands of unrelated network packets
    \end{itemize}
\item Full Network Topology
    \begin{itemize}
        \item Use as reference for identities of specific devices, current security architecture (flat network), likely paths for lateral traversal, etc.
        \item The agent is a worker so they will have access to all materials that the real SOC would
        \item Trap: Include several unrelated devices to obscure relevant IP devices to the attack
    \end{itemize}
\item Operating System (OS) Logs
    \begin{itemize}
        \item Identify malicious host-level activity, such as unusual process creation, file encryption, etc.
        \item We provide logs for both the Microsoft Windows and Linux operating systems. (\textbf{Microsoft Windows:} Windows Events, Microsoft Defender, \textbf{Linux:} Syslog, Systemd)
        \item Trap: Include system noise through large volume of benign user actions.
    \end{itemize}
\item Firewall Logs
    \begin{itemize}
        \item Identify malicious network activity, such as port scanning, failed SSH attempt, etc.
        \item We provide logs for both the Microsoft Windows and Linux operating systems. (\textbf{Microsoft Windows:} Microsoft Defender for Endpoint, \textbf{Linux:} UFW, firewalld)
        \item Trap: Include miscellaneous network traffic and blocked connections to obscure attack patterns.
    \end{itemize}
\item SIEM Logs
    \begin{itemize}
        \item Identify malicious activity across hosts and the network through rule-based correlation, severity labeling, etc.
        \item We provide logs for contemporary SIEM platforms. (Splunk Enterprise Security, Microsoft Sentinel)
        \item Trap: Include system noise through overly broad SIEM rules.
    \end{itemize}
\item XDR Alerts
    \begin{itemize}
        \item Identify malicious activity across hosts and the network through anomaly-based correlation, machine learning metrics, etc.
        \item XDR alert systems are an increasingly common component of the SOC environment. Task Tiger includes \textbf{XDR Alerts} as a forward-looking data source.
        \item We provide alert data for contemporary XDR platforms. (Microsoft Defender XDR, CrowdStrike Falcon)
        \item Trap: Include system noise through frequent false positive alerts.
    \end{itemize}
\item Management Service Logs
    \begin{itemize}
        \item Identify malicious identity and administrative activity, such as privilege escalation and unauthorized account changes.
        \item We provide logs for Microsoft management services.   (Microsoft Active Directory)
        \item Trap: Include frequent password resets and account lockouts to obscure credential abuse.
    \end{itemize}
\item Malware Samples
    \begin{itemize}
        \item Identify malware activity based on binary analysis output, network calls, name of executable, etc.
        \item Agent can use information from solving Task Fox as evidence for Task Tiger
        \item Trap: Include unrelated, latent malware samples (e.g., many employees download tools from poisoned websites)
    \end{itemize}
\item Disk Images
    \begin{itemize}
        \item Identify targets of malicious activity based on filesystem access, modified disk permissions, data exfiltration, etc.
        \item Agent can use information from solving Task Goat as evidence for Task Tiger
        \item Trap: Introduce real-world complexity with messy directory structures, symbolically linked across multiple hard drives.
    \end{itemize}
\item User Logs
    \begin{itemize}
        \item Identify malicious activity based on frequent user complaints. To follow \textit{DP3}, \textbf{User Logs} will follow a timed release and only become available at some point during the second half of SOC-bench. In real-world incidents, users only notice that something is wrong much later than security systems.
        \item Trap: Upset users can write unclear tickets that accidentally mislead SOC analysts.
    \end{itemize}
\end{enumerate}

\subsection{Scoreboard Design}
The scoreboard for Task Tiger is measured through three categories of decision-making in the role of a Tier 2 SOC analyst. Each metric will be applied against a sequence of events observed from the Colonial Pipeline incident. The agent will be required to 1. identify relevant data sources for an incident response report, 2. build an accurate threat graph for the sequence of events from initial entry to the current state of the attack, and 3. summarize the specific, initial entrypoint used to initiate the large-scale attack. Agents will not receive immediate feedback for submissions. Scored feedback will only be returned during each of the standard 30-minute intervals. This ensures loyalty to \textit{DP1} of SOC-bench, where senior analysts are encouraged to create a few correct reports rather than frequently updating an incorrect report. 

\begin{enumerate}
    \item Relevant Data Sources
        \begin{itemize}
            \item An introductory category for agentic decision-making in Task Tiger: Agents must identify which of the available data source are indicative of characteristics from the large-scale attack. Not all data sources provided in Task Tiger will contain relevant logs or other information. This is intended to stay loyal to \textit{DP1} of SOC-bench, as a real SOC will process massive amounts of IT data that is irrelevant to an incident. 
            \item It is up to the agent to determine why a data source may warrant further investigation. Accurate decision-making will both benefit the agent's score and assist with pursuing the correct information for subsequent outcomes. During early scoring intervals, this should be considered as a preliminary report. Once all outcomes are satisfied, the agent should consider this to be an accurate introduction to the full incident response report.
            \item To further evaluate the agent's understanding as a Tier 2 SOC analyst, we will require the agent to denote when two or more data sources have a \textbf{relationship}. Related data sources may contain characteristics like log events that coincide with each other. Such connections will contribute towards understanding the progression of events during the large-scale attack.
            \item Incorrect actions will result in a small deduction.
        \end{itemize}
    \item Threat Graph
        \begin{itemize}
            \item A secondary category for agentic decision-making in Task Tiger: Agents must recreate the sequence of events in the large-scale attack from initial entry until the current state during the scoring interval. This category requires agents to classify malicious activity discovered in the \textbf{Relevant Data Sources}. This is intended to challenge agents against a more structured, data-driven approach to decision-making.
            \item To remain loyal to \textit{DP2} of SOC-bench, the agent must follow the format of a threat graph. This is an industry standard tool in multiple cybersecurity vendors (e.g., Crowdstrike, Splunk), allowing Tier 2 analysts to document current knowledge on an attack incident.
            \item In the threat graph format, a specific artifact (e.g., IP addresses, hostnames, malicious files) denotes a node. An event relationship between two or more artifacts (e.g., A server 192.168.10.2 uploading a file to a workstation 192.168.20.5) denotes an edge. Not every node will necessarily have a parent/child edge attached to it.
            \item Agents will also be required to denote whether a given artifact suffered from a significant vulnerability. Like with real-world incidents, certain security flaws must be highlighted as most responsible for initiating, then escalating the large-scale attack.
            \item The \textbf{User Support Ticket} data source will only be released after several scoring intervals have passed. There will be deductions for referencing such tickets as evidence. An agent relying on support tickets is using a data source that indicates the large-scale attack was already successful.
            \item Incorrect actions will result in a small deduction.
        \end{itemize}
    \item Initial Entrypoint    
        \begin{itemize}
            \item A final category for agentic decision-making in Task Tiger: Agents must produce an executive summary of the findings as a Tier 2 SOC analyst, including confirmation of the initial entrypoint. This category demonstrates that the agent was successful in its technical investigation. Additionally, it shows that the agent understands how to prioritize and present its information. As a Tier 2 SOC analyst, the agent is responsible for both technical and written outcomes.
            \item This executive summary will also require evidence to back up assertions from the agent. There should be specific references to findings inside specific data sources.
            \item Incorrect actions will result in a small deduction.
        \end{itemize}
    % \item Bonus
    %     \begin{itemize}
    %         \item A tertiary category for achieving points in Task Tiger. As a forward-looking benchmark, the following metrics are optional and evaluate how an AI can improve Tier 2 SOC performance.
    %     \end{itemize}
    %     \begin{table}[H]
    %     \centering
    %     \begin{tabularx}{\linewidth}{l|X|r}
    %     Metric & Description & Points \\\hline
    %     Efficiency & Correctly completes Task Tiger within an early stage ( every 30 minutes). & $1 \: point \cdot \frac{Maximum \: Rounds}{Rounds \: To \: Complete}$\\
    %     \end{tabularx}
    %     \caption{\label{tab:widgets}Bonus Rubric.}
    %     \end{table}
\end{enumerate}
\fbox{%
  \parbox{\textwidth}{%
    \textbf{Note:} In the context of the Task Tiger Scoreboard, the term \textit{each} refers to a deduction for each instance of the action.
  }%
}
\begin{table}[H]
\centering
\begin{tabularx}{\linewidth}{@{}lXc@{}}
\toprule
\textbf{Outcome/Action} & \textbf{Description} & \textbf{Available Points} \\ 
\midrule
\textbf{Relevant Data Sources} & & 15 points total \\
Irrelevant Data Source & Identifies a data source that is not part of the large-scale attack.  & -2 points each \\
Inaccurate Relationship & Incorrectly identifies that information correlates between two or more data sources. & -1 points each \\
\midrule
\textbf{Threat Graph} & & 35 points total \\
\textit{Node Accuracy} & & \\
Incorrect Artifact & Identifies artifact (i.e., IP Address, Process ID) unrelated to large-scale attack. &  -1 point each \\
Irrelevant Vulnerability & Identifies security flaw that does not affect the given artifact. &  -1 point each \\
Incorrect Source Attribution & References a \textbf{Relevant Data Source} that does not support the specific artifact. &  -1 point each \\
Support Ticket Attribution & References \textbf{User Support Tickets} as supporting data source, indicating delayed action. & -5 points \\
\textit{Edge Accuracy} & & \\
Incorrect Relationship & Incorrectly identifies a relationship between two nodes that have not interacted. &  -1 point each \\
Incorrect Description & Describes an interaction event (i.e., Malicious process began) that did not occur between two nodes. &  -1 point each \\
Incorrect Source Attribution & References a \textbf{Relevant Data Source} that does not support the specific interaction. &  -1 point each \\
Support Ticket Attribution & References \textbf{User Support Tickets} as supporting data source, indicating delayed action. & -5 points \\
\midrule
\textbf{Initial Entrypoint} & & 50 points total \\
Incorrect Entrypoint & Identifies the wrong entrypoint within the summary. & -25 points \\
Insufficient Evidence & Does not provide sufficient evidence referencing data sources. & -25 points \\
\bottomrule
\end{tabularx}
\caption{Scoreboard Rubric.}
\end{table}

\subsection{Reference: Ground Truth Threat Graph}

\fbox{%
  \parbox{\textwidth}{%
    \textbf{Warning:} AI agents are not permitted to read this section of content.
  }%
}

Task Tiger is observed from a real-world, large-scale attack campaign referred to as the \textbf{Colonial Pipeline Incident}. In this incident, the \textbf{Colonial Pipeline Company} suffered from a targeted attack by a single advanced persistent threat (APT) group. Task Tiger is faithful to the design principles of SOC-bench, and therefore, it attempts to follow this structure closely. Significant events throughout the attack campaign should be tightly correlated with the specialties of the APT group. The following figure is a sample to illustrate how an industry-standard threat graph may be structured for this scenario. This approach honors \textit{DP3} of SOC-bench, ensuring that agents are evaluated with a realistic threat graph as ground truth.

\begin{figure}[H]
\includegraphics[width=\textwidth]{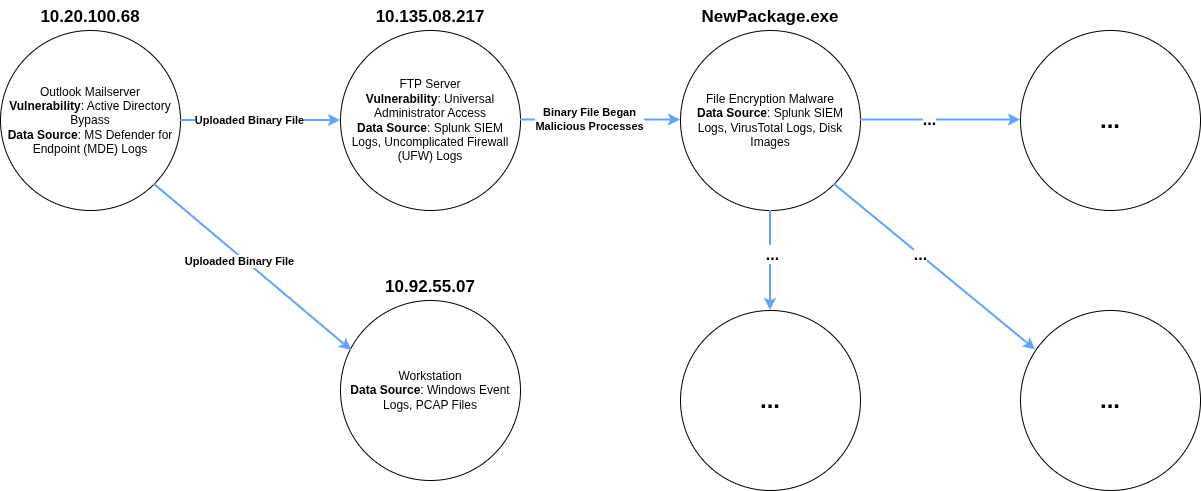}
\caption{Sample of Ground Truth Threat Graph}
\end{figure}
\noindent
\subsection{Remarks}
\begin{enumerate}
    \item Accessibility of Threat Graphs
    \begin{itemize}
        \item Following each of the design principles for SOC-bench, we wish to challenge agentic AI frameworks with real-world security responsibilities. In security analysis, many SIEM and XDR vendors provide cutting-edge tools for visualizing incidents as a threat graph. However, these tools are often closed-source and few companies publish datasets.
    \end{itemize}
    \item Non-deterministic Nature of the Tier 2 SOC
    \begin{itemize}
        \item Within the scope of SOC-bench, Task Tiger is distinct from other tasks with its focus on Tier 2 of the SOC. This can incorporate unique complexities, as Tier 2 SOC responsibilities are much more open-ended. Whereas a Tier 1 analyst needs to follow a well-defined set of rules, Tier 2 analysts are tasked with more vague, investigative tasks. This role incorporates personal opinion based on years of experience to deliver results. Contemporary AI agents may struggle with this aspect, but this means that Task Tiger will provide fascinating insight towards their thought process for SOC decision-making. 
    \end{itemize}
\end{enumerate}

%-----------------------
\section{Task Panda}
\label{sec:taskpanda}

\subsection{Task Overview}

Task Panda evaluates a multi-agent AI system’s ability to reason over numerous data sources and generate timely and accurate containment responses during a Colonial Pipeline enterprise ransomware incident \cite{cisa_colonial_pipeline_lessons_2023,CISA-AA21-131A,doe_colonial_pipeline_2021,nist_sp800_61r3_2025}. In line with design principle DP2, this task focuses solely on containment actions, without considering other incident response phases \cite{nist_sp800_61r3_2025}.

Following design principle DP4, the scenario unfolds along a timeline divided into 30-minute stages. At each stage, the AI system receives a variety of data sources reflecting system activity, network traffic, performance metrics, and threat intelligence. These data sources are provided to simulate the data sources available to a real SOC, in accordance with DP1 \cite{nist_sp800_61r3_2025,mitre_attack_data_sources}. Detailed descriptions of these data sources are provided in the Data Sources section.

At each stage, the AI system must synthesize information from the data sources to determine whether and how to act effectively and produce a BLUF (Bottom Line Up Front) report summarizing the containment actions, affected assets, supporting evidence and reasoning \cite{nist_sp800_61r3_2025}. Reporting requirements and formatting are detailed in the Outcomes section.

For every stage, SOC Bench defines the ground truth representing the optimal SOC response. A scoreboard evaluates the AI system’s accuracy, completeness and reasoning of containment actions based on the submitted BLUF reports \cite{mitre_engenuity_methodology_overview,zhang2025cybench,wang2026cybergym}.

\ReaderDisclaimer{Section \ref{sec:taskpanda}}{This task introduces inherent inter-task dependencies to test the AI system’s reasoning capabilities; however, following design principle DP2, such dependencies are not supposed to be disclosed to any AI system that is evaluated against SOC-bench.} 

%---------------------
\subsection{\textbf{}\textbf{Data Sources}}

All data sources provided to the AI system simulate the data sources available to a real SOC during the ransomware incident. In accordance with design principles DP1 and DP3, the data strictly follows the Colonial Pipeline ransomware incident, reflecting the actual attack process and operational context. These sources are time-stamped and synchronized to provide realistic evidence for containment decision-making.

\subsubsection{Filesystem Operations}

Records of file modifications, creations, deletions, and encryption activity on monitored endpoints and storage systems. The corresponding logs provide insight into which files are impacted and the progression of ransomware activity.

\subsubsection{Network Traffic}

The corresponding PCAP files reveal communication patterns, lateral movement, command-and-control activity, and abnormal network behaviors.

\subsubsection{Operating System Logs}

Operating system logs provide visibility into authentication activity, process execution, service operations, and overall system behavior. Windows Event Logs capture security events such as logon attempts, process creation, PowerShell execution, and service or registry changes. Linux journal logs (journald) record system messages, authentication attempts, privilege escalations, service restarts, and kernel or daemon alerts. These logs help correlate host-level activity with ransomware behavior and support evidence-based containment decisions.

\subsubsection{Host-level Measurements}

Time-series measurements of CPU usage, memory consumption, I/O activity, and network throughput. Spikes or deviations can indicate ransomware operations or other anomalous behavior on hosts.

\subsubsection{Cyber Threat Intelligence (CTI)}

CTIs are Structured reports including indicators of compromise (IOCs), tactics, techniques, and procedures (TTPs) relevant to ransomware campaigns. CTI provides contextual knowledge for interpreting observed events and supporting containment decisions.

\subsubsection{Network Topology Manifest}

A structural mapping of hosts, VLANs, segments, and firewall zones in the simulated environment. This allows the AI system to understand containment boundaries and logical relationships between assets.

\subsection{Outcomes}
\label{sec:pandaoutcome}

The AI system may produce a BLUF report at each stage; this report is the only evaluable output of Task Panda. It captures the agent’s containment actions, affected assets, and the reasoning supporting these decisions. All expected outcomes follow the real-world Colonial Pipeline ransomware incident, in accordance with design principle DP3.

If the agent determines that no containment action is necessary, it may still submit a BLUF report that omits actions and targets but provides reasoning and evidence explaining why inaction is justified. This mirrors real SOC decision-making, where inaction can be the correct and evidence-based choice when the available data does not support initiating containment.

\subsubsection{BLUF Report Content Requirements}

Every BLUF report must be structured in JSON format and must include the fields below. Each field is evaluated directly on the scoreboard.

\textbf{(1) }\textbf{Stage timestamp}

The end time of the stage when the decision is made.

\textbf{(2) }\textbf{Containment actions}

A list of all containment actions the agent proposes for this stage. Multiple actions may be included (e.g., isolations, firewall blocks, credential resets).

\textbf{(3) }\textbf{Action targets}

A structured list of all targets affected by the containment actions. Since real containment may involve many assets, this field supports multiple categories:

\begin{itemize}
    \item hosts: endpoint hostnames or IDs
    \item ips: individual IP addresses
    \item subnets: network segments or VLANs
    \item user\_accounts: individual compromised accounts
    \item user\_groups: AD or IAM groups
    \item services: specific processes or enterprise services
    \item other: any additional SOC-relevant categories
\end{itemize}
Agents must list only assets directly affected by their containment actions.

\textbf{(4) }\textbf{Reasoning and evidence}

A concise BLUF-style justification that integrates three required elements:

(I) Action Justification Based on SOC Trade-Offs

The agent must explain why the selected containment actions are appropriate considering the trade-offs between:

\begin{itemize}
    \item data security impact
    \item availability impact
    \item business/operational continuity
\end{itemize}
(II) assessment of Current Situation

The reasoning must assess the agent’s evaluation of current situation based on data sources, such as:

\begin{itemize}
    \item encryption activity
    \item lateral movement
    \item C2 connections
    \item exfiltration attempts
    \item compromised accounts
    \item anomalous host or network behavior
\end{itemize}
The assessment report must include concrete evidences (e.g. log records). This assessment must support the chosen actions and targets.

(III) Predicted Impact if No Action Is Taken

A brief forward-looking statement describing likely consequences of inaction, such as:

\begin{itemize}
    \item ransomware spreading to adjacent hosts
    \item privilege escalation
    \item loss of business systems
    \item increased data exfiltration
    \item service outages
    \item degraded operational technology (OT) availability
\end{itemize}
This aligns BLUF reasoning with real SOC decision-making.

\subsubsection{BLUF Report Format Requirements}

All BLUF reports must follow standardized JSON formatting to ensure consistent scoring. Each report must be:

\begin{itemize}
    \item structured (JSON only)
    \item self-contained
    \item time stamped
    \item explicitly listing actions, targets, and reasoning
\end{itemize}
Example JSON-formatted BLUF Report:
\begin{table}[h]
\centering

\begin{tabular}{|>{\raggedright\arraybackslash}p{1\linewidth}|}
\hline
\{

"stage\_timestamp": "01/01/2025, 00:00:00 AM",

"containment\_actions": [

"Isolate ENG-WKS-12 and ENG-WKS-17 from the corporate LAN",

"Block outbound SMB and HTTPS traffic from ENG-SRV-05 to 185.199.110.0/24"

],

"action\_targets": \{

"hosts": ["ENG-WKS-12", "ENG-WKS-17", "ENG-SRV-05"],

"ips": [],

"subnets": ["185.199.110.0/24"],

"user\_accounts": [],

"user\_groups": [],

"services": [],

"other": []

\},

"reasoning\_evidence": "Isolation and outbound blocking are selected to maximize security impact while minimizing operational disruption: both workstations are non-critical endpoints and blocking ENG-SRV-05 outbound traffic preserves service availability while halting lateral movement. Current data shows ENG-WKS-12 and ENG-WKS-17 exhibit rapid encryption and unauthorized C2 connections. ENG-SRV-05 is initiating anomalous SMB flows to a malicious subnet. If containment is delayed, ransomware will likely propagate across engineering hosts and enable further data exfiltration through ENG-SRV-05."

\} \\
\hline

\end{tabular}

\end{table}
\subsection{Scoreboard}

The AI system’s performance is evaluated per stage, based on the submitted BLUF reports. The scoreboard measures the accuracy, completeness, and quality of reasoning for containment decisions.

\subsubsection{Scoreboard Components}

 \textbf{(1) Containment actions—Maximum 20 points}

Evaluates whether the containment actions match the corresponding stage’s ground truth defined by SOC Bench.

\begin{table}[h]
\centering

\begin{tabular}{|>{\raggedright\arraybackslash}p{0.3\linewidth}|>{\raggedright\arraybackslash}p{0.1\linewidth}|>{\raggedright\arraybackslash}p{0.5\linewidth}|}
\hline
Level & Points & Description \\
\hline
Fully correct & 20 & All actions exactly match the ground truth. \\
\hline
Partially correct & 10 & Actions address part of the ground truth but are incomplete or partially misapplied. \\
\hline
Incorrect or missing & -20& Actions are completely wrong, absent, or mostly incorrect, increasing operational risk and potential consequences. \\
\hline

\end{tabular}

\end{table}

\textbf{(2) Action Targets—Maximum 40 points}

Evaluates whether the agent correctly identifies all assets affected by the containment actions.

\begin{tabular}{|>{\centering\arraybackslash}p{0.3\linewidth}|>{\centering\arraybackslash}p{0.1\linewidth}|>{\raggedright\arraybackslash}p{0.5\linewidth}|}
\hline
Level & Points & Description \\
\hline
Fully correct & 40 & Every affected asset is correctly identified. \\
\hline
Partially correct & 20 & Some assets are correctly identified, but key targets are missing or extra irrelevant assets included. \\
\hline
Incorrect or missing & -20& Most or all targets are incorrect or missing, which would likely cause ineffective containment or considerable collateral impact. \\
\hline

\end{tabular}

\textbf{(3) Reasoning and Evidence—Maximum 40 points}

Evaluates whether the agent provides coherent reasoning as described in Section \ref{sec:pandaoutcome}.

\begin{table}[h]
\centering

\begin{tabular}{|>{\raggedright\arraybackslash}p{0.3\linewidth}|>{\raggedright\arraybackslash}p{0.1\linewidth}|>{\raggedright\arraybackslash}p{0.5\linewidth}|}
\hline
Level & Points & Description \\
\hline
Fully supported & 40 & Reasoning is clear, logical, and fully supported by evidence from the provided data sources. \\
\hline
Partially supported & 20 & Reasoning is somewhat clear but incomplete or partially unsupported by evidence. \\
\hline
Unclear or unsupported & -20 & Reasoning is unclear, illogical, or unsupported. \\
\hline

\end{tabular}

\end{table}

\begin{tightitemize}
  \item “No-action” BLUF reports: If the agent submits a report without containment actions or targets, these fields are scored according to the ground truth (which may itself prescribe “no action” and “no target”), and Reasoning and Evidence are evaluated accordingly.

  \vspace{1em} % adds an empty line

  \item “No-action” BLUF reports: If the agent submits a report without containment actions or targets, these fields are scored according to the ground truth (which may itself prescribe “no action” and “no target”), and Reasoning and Evidence are evaluated accordingly.
\end{tightitemize}

\ReaderDisclaimer{Scoreboard}{SOC-bench embeds inherent inter-task dependencies to test the AI system’s reasoning capabilities. Following design principle DP2, these dependencies are not supposed to be disclosed to the AI system or developers.}

\subsubsection{Score Calculation}

Since the total number of stages is fixed, the Final Score is computed as the sum of per-stage scores:

Where: 

StageScore\_i = ContainmentDecision + ActionTargets + ReasoningAndEvidence + TrapPenalty

\begin{itemize}
    \item N = the number of evaluation stages
    \item Range per stage: -80 to 100 (including trap penalties)
    \item Final Score range: -80 * N to 100 * N
\end{itemize}

\subsection{Remarks}

Task Panda evaluates AI systems per stage, with stages occurring every 30 minutes. Each stage has a ground truth used for scoring.

Effective containment requires correlating multiple independent data sources to identify valid actions.

Timeliness is inherently reflected in scoring: accurate and well-reasoned containment decisions submitted in a timely manner are rewarded. Premature actions without sufficient evidence or delayed responses that miss the stage window are implicitly penalized via negative scores.

All data sources are timestamped and synchronized to 30-minute stage to ensure consistent temporal alignment with real SOC operations, which is in keeping with DP1.

%\ReaderDisclaimer{Remarks}{The inherent inter-task dependencies indicate that individual agents could suffer a lot if they do not coordinate with other agents; coordination is often required to avoid subtle misleading indicators. The dependencies are not supposed to be disclosed to the AI system or its developers.}

\subsection{Challenges}

\ReaderDisclaimer{Challenges}{The content in this section is not supposed to be disclosed to any AI system evaluated against SOC-bench.} 

Designing Task Panda involves several challenges that directly affect the benchmark’s realism, fairness, and reproducibility. The following key issues define the core difficulties.

\subsubsection{Managing Incomplete and Asynchronous Evidence}

In accordance with DP1, evidence is provided incrementally to reflect real SOC data collection. Evidence from key sources may not be available at specific evaluation stage. The design balances sufficiency against investigative uncertainty: too little data prevents reasoning, while too much reduces realism.

\subsubsection{Defining Dynamic and Valid Ground Truth}

Stage containment actions are often situational rather than absolute. Multiple actions can be valid depending on evidence, operational priorities, or containment scope. Establishing canonical “correct” actions requires expert synthesis to ensure ground truth is consistent, defensible, and adaptable across scenarios.

\section{Conclusion}
%Your conclusion here

The goal of this work is to develop a set of design principles for the construction of a benchmark, which is denoted as SOC-bench, to evaluate blue team capabilities of AI systems \& agents. 
In particular, five design principles are identified: (DP1) Except for only allowing humans to serve as a supervisor, the as-is SOC (incident response) system, 
not the to-be SOC system, serves as the gold standard. 
(DP2) Because AI systems should autonomously explore and leverage the interdependence between the
family of blue team tasks, SOC-bench should not provide AI systems with any cross-task hints. 
(DP3) AI systems could go beyond mimicking humans.
(DP4) Real-world SOC (incident response) systems are inherently imperfect.
(DP5) SOC-bench should not be specific to any current capabilities of AI systems. 
In order to validate these design principles, we have followed these principles and developed a conceptual design of SOC-bench, which consists of a family of five blue team tasks in the context of large-scale ransomware attack incident response: (task Fox) Early cyberattack campaign detection; (task Goat) file system forensics; (task Mouse) data exfiltration analysis; (task Tiger) analysis of IOCs to attribute the attack and report the used TTPs; (task Panda) recommending containment actions. 

% \section*{Acknowledgments}
% This was was supported in part by......

%Bibliography
\bibliographystyle{unsrt}  
\bibliography{references} 
\end{document}